# Deep DNA Storage: Scalable and Robust DNA-based Storage via Coding Theory and Deep Learning

Daniella Bar-Lev*, Itai Orr*, Omer Sabary*, Tuvi Etzion and Eitan Yaakobi
Department of Computer Science, Technion - Israel Institute of Technology, Haifa, Israel
*These authors contributed equally

## Abstract
DNA-based storage is an emerging technology that enables digital information to be archived in DNA molecules. This method enjoys major advantages over magnetic and optical storage solutions such as exceptional information density, enhanced data durability, and negligible power consumption to maintain data integrity[1,2]. To access the data, an information retrieval process is employed, where some of the main bottlenecks are the scalability and accuracy, which have a natural tradeoff between the two. Here we show a modular and holistic approach that combines Deep Neural Networks (DNN) trained on simulated data, Tensor-Product (TP) based Error-Correcting Codes (ECC), and a safety margin mechanism into a single coherent pipeline. We demonstrated our solution on 3.1MB of information using two different sequencing technologies. Our work improves upon the current leading solutions by up to x3200 increase in speed, 40% improvement in accuracy, and offers a code rate of 1.6 bits per base in a high noise regime. In a broader sense, our work shows a viable path to commercial DNA storage solutions hindered by current information retrieval processes.

## DNA-based storage
There is an exponential growth in the global data sphere fueled by the proliferation of digital technologies such as artificial intelligence, the Internet of Things, widespread internet connectivity, and the growing number of interconnected devices. While the global data sphere is anticipated to reach 180 Zettabytes by 2025, current storage solutions are not expected to scale at nearly the same pace due to capacity limitations[3]. To address this urgent need of the digital age, researchers are turning to innovative solutions like DNA-based storage, recognizing its potential to revolutionize long-term data storage capabilities due to its extraordinary data density and durability[2].
A DNA molecule consists of four building blocks called nucleotides: Adenine (A), Cytosine (C), Guanine (G), and Thymine (T). A single DNA strand, also called oligonucleotide, is an ordered sequence of some combination of these nucleotides and can be abstracted as a string over the alphabet $\{A, C, G, T\}$. The ability to chemically synthesize almost any possible nucleotides sequence makes it possible to store digital data on DNA strands.
The standard in-vitro DNA-based storage pipeline consists of several steps and is shown in Fig. 1a. First, the binary data is encoded into sequences over the DNA 4-ary alphabet, which are referred to as encoded sequences. Next, the encoded sequences are synthesized by a DNA synthesizer. The DNA synthesizer produces multiple DNA strands (known as oligos) for each encoded sequence, as current synthesis technologies cannot produce one single strand per sequence. Moreover, the length of the strands produced by the synthesizer is typically bounded by roughly 200-300 nucleotides to sustain an acceptable error rate[4]. The synthesized strands are then stored in a storage container in an unordered manner. To access the data, a sample of the strands is taken from the storage container, amplified using Polymerase Chain Reaction (PCR), and then sequenced by a DNA sequencer. The sequencer processes the strands and generates reads, which are digital representations of the strands as sequences over the DNA alphabet. However, at this stage, the data is mixed in an unordered and random manner.

Recovery of the original binary information is then obtained by a computational-heavy step, which is referred to as DNA information retrieval.
DNA as a storage medium has several unique attributes that distinguish it from its widespread digital counterparts and should be considered in the design of the information retrieval pipeline. The first attribute is the inherent redundancy obtained by the synthesis and the sequencing processes, where each synthesized DNA strand has several copies. The second is that the strands are not ordered in memory and thus it is not possible to know the order in which they were stored. The third attribute is the unique noise characteristics from both the synthesis and the sequencing processes which introduce errors to the reads. These errors affect data integrity and are mostly of three types, insertion, deletion, and substitution of symbols, where the error rates and their characteristics depend on the synthesis and sequencing technologies[5].
The information retrieval pipeline, also shown in Fig. 1a, can be partitioned into several main stages: clustering, reconstruction, and decoding. First, a clustering algorithm is performed on the obtained reads. In this step, the unordered set of reads is partitioned into groups, where the goal is to partition the reads such that all the reads in each group originate from the same encoded sequence. Second, a reconstruction algorithm should be applied to each cluster to predict the encoded sequence. The use of a clustering algorithm and a reconstruction algorithm utilizes the inherent redundancy of DNA synthesis and sequencing to correct most of the errors, but usually not all.
In the DNA reconstruction problem[6–8], the goal is to predict an encoded sequence from a set of reads. One of the challenges in the DNA-based storage channel is that we do not necessarily have control over the size of a cluster, and it is likely that this size is significantly smaller than the required minimum size that guarantees a successful reconstruction of the encoded sequence according to the classical Levenshtein reconstruction problem[9]. Lastly, the decoding step converts the reconstructed encoded sequence back to the binary data. In this step, if an ECC was applied during the encoding phase (prior to the DNA synthesis step), the remaining errors can be corrected using the decoding procedure of the ECC.
The first large-scale experiments that demonstrated the potential of in vitro DNA storage were reported by Church et al.[10] who stored 643KB of data and Goldman et al.[11] who accomplished the same task for a 739KB message. However, neither group recovered the entire message successfully. Later, Grass et al.[12] used a Reed Solomon (RS) based coding[13] solution in their DNA-based storage experiment, where they stored and recovered successfully an 81KB message. Erlich and Zielinski presented DNA Fountain[14], a coding scheme that is based on Luby transform. In their experiment, they stored and recovered 2.11MB of data. Organick et al.[15] developed and demonstrated a scheme that allows random access using DNA strands, where they stored 200MB of data. Yazdi et al.[16] presented a new coding scheme that was designed to correct errors from Nanopore sequencers, a smaller and portable sequencing technology that allows longer strands but has higher error rates. In their experiment they encoded 3.633KB of data which was successfully recovered. Wang et al. [17] stored a 379.1KB using their suggested inner-outer scheme that combines cyclic redundancy check and repeat accumulate code while approaching an information capacity of 1.69 bits per base. Chandak et al.[18] suggested a coding method based on convolutional code and Recurrent Neural Network that integrates with Nanopore MinION. They were able to demonstrate their method and stored 11KB of information. Anavy et al.[19] demonstrated how the capacity of the DNA-based storage can be increased using composite letters.
The design of an information retrieval pipeline for DNA-based storage is a challenging problem as the errors include deletions and insertions. These errors are challenging types of errors, and many of the related theoretical problems are far from being solved[20]. This fact makes the design of both clustering and reconstruction algorithms more complex[21,22]. Additionally, the clustering problem is a computationally intensive problem by itself. This is especially challenging when applied to DNA-based storage where the number of clusters can be extremely large, for example, 1TB of data will require an order of billions of clusters. Furthermore, theoretical reconstruction algorithms designed to address deletion and insertion errors, usually assumed that the clusters were partitioned (almost) perfectly[6,23,24] or were designed to work on a large block-length[23,25–28].

Several works[12,14,18] tackled these difficulties by adding redundant symbols to each designed DNA strand (i.e., inner coding), or by adding redundant DNA strands (i.e., outer coding) to detect and correct the deletion and insertion errors. In these techniques, the clustering and the reconstruction steps can be avoided. In this approach, the inherent redundancy of the synthesis and the sequencing processes is not fully utilized which leads to suboptimal use of redundancy in the design of the ECC. This in turn can also lead to performance degradation due to an increase in the number of strands (and therefore also reads) that represent the information and should be processed.

Using machine learning methods for DNA-based storage was examined in Bee et al.[29], where the authors proposed a content-based similarity search and demonstrated how it can be added to DNA-based storage systems. In Pan et al.[30], the authors showed how data can be stored in DNA both in the strand and in its backbone structure, to create a rewriteable DNA-based storage system. Since these methods rely on the low entropy of the data, they are mainly suitable in cases where there is a structured pattern in the data that the reconstruction method can exploit.

## End-to-end solution for DNA information retrieval

In this work, we present an end-to-end, practical solution to the in-vitro DNA information retrieval problem, as shown in Fig. 1a. Our solution, termed DNAformer, utilizes a modular encoding scheme, combining ECC and constrained codes prior to DNA synthesis and storage. Our coding scheme enables the pragmatic partitioning of large datasets into smaller blocks to allow for fast and easy access to specific parts within the data. When the information requires access, a sequencing process is used followed by an information retrieval process. The first step in this process is to bin the different reads into groups based on their index. This naïve approach introduces noise into the clusters which we treat in the following steps. The benefit is a significant increase in the speed of the clustering step over alternatives, more noise-tolerant approaches, such as the hash-based approach[22] and the Clover clustering approach[21].

In the second step, we utilize a DNN to reconstruct a sequence over the DNA alphabet based on a non-fixed number of reads with varying lengths and solve a sequence-to-sequence, Multiple-In-Single-Out problem. The model architecture shown in Fig. 1b uses a combination of convolutions and transformers followed by a confidence filter to screen correct predictions from false ones. The suspected incorrect predictions can go through a second reconstruction step, a dynamic programming-based algorithm, termed Conditional Probability Logic (CPL). The CPL algorithm analyzes the reads in each cluster and estimates their possible errors and probabilities according to the reads' similarity in the edit distance metric. The main advantage of the CPL algorithm is that it requires zero prior-knowledge of the cluster's error probabilities, and by estimating them it is possible to predict a good estimation even for small and erroneous clusters. This step adds another degree of freedom in our solution to deal with the tradeoff between accuracy and speed where the purpose is to deal with more challenging cases while not sacrificing the run-time capabilities of the system. Furthermore, our approach uses the concept of safety margin to quantify how robust the information retrieval pipeline is under specific working conditions.

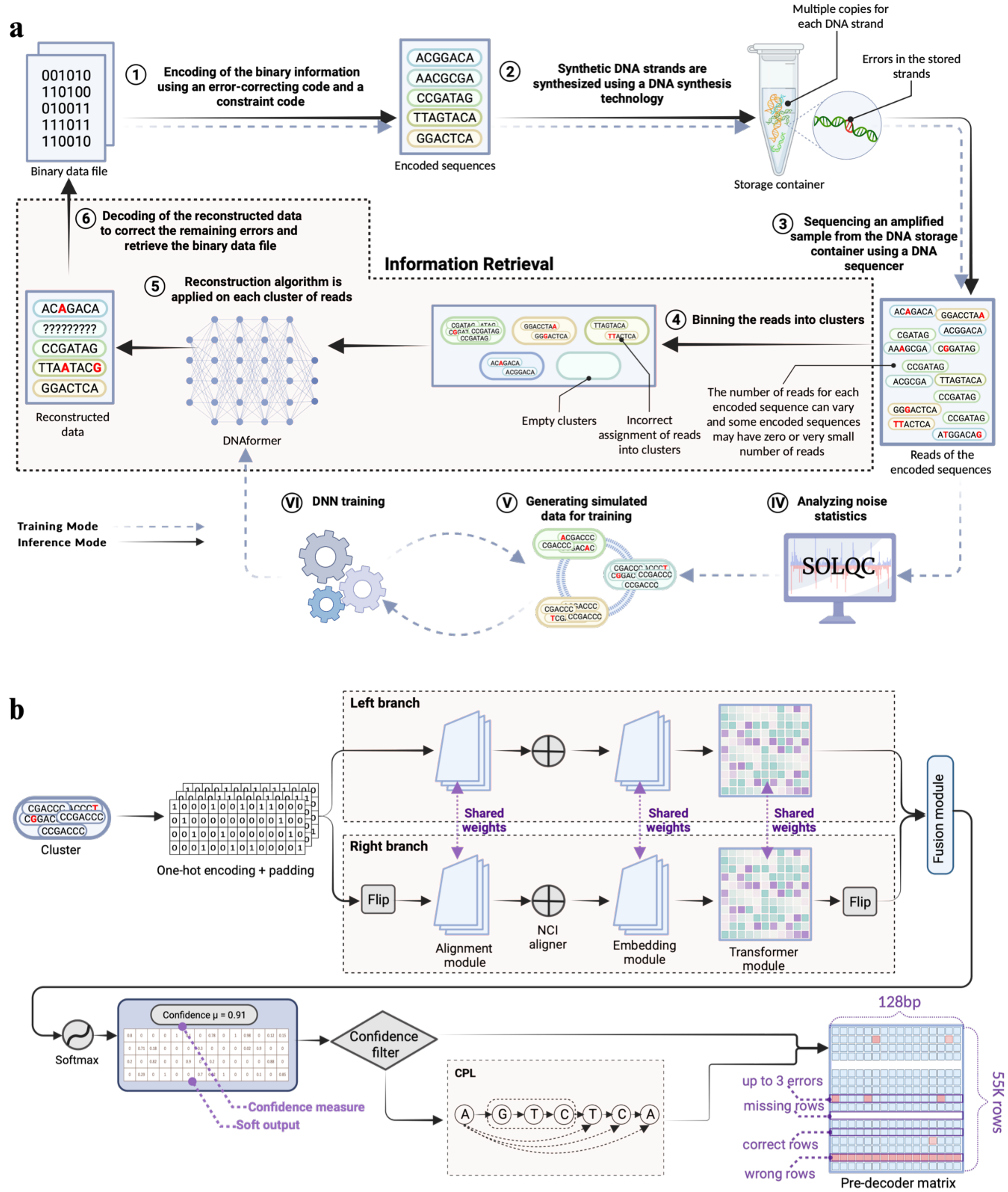

**Fig. 1 | End-to-end solution for DNA information retrieval. a,** shows a schematic description of our solution for the DNA-based storage pipeline. The numbers 1-6 depict the different stages through the process. In Roman numerals**,** we depict steps that are part of the training pipeline. **b,** shows a detailed view of the information retrieval process showing the DNN architecture, confidence filer, CPL, and the input to the decoder.

The third step is the decoding of the data, where we utilize a modified TP code[31]. In our solution we modify the standard TP approach to take advantage of the first two steps, the clustering and reconstruction steps, and create a coherent, unified pipeline. This approach allows us to take advantage of the inherent redundancy and success of the upstream steps to reduce the redundancy within the coding scheme. Our scheme allows simple integration with a large family of constrained codes and flexibility with error correction capabilities.
Due to the high cost of acquiring large amounts of real data sufficient for training a DNN, we based our approach on simulated data. Our training methodology uses only a small amount of real data to model the error rates during the synthesis, PCR, and sequencing processes, and is done using the SOLQC tool[32]. Once these errors were modeled, simulated data can be generated to train a DNN in any quantity required. An important distinguishing property of this methodology is that the errors need to be modeled only once for each synthesis and sequencing processes, which makes our method scalable and cost effective.
A key point of our solution strategy is that we do not teach the model to utilize underlying semantics and file structure, but rather focus on the noise characteristics of the synthesis and sequencing processes. This gives the model the important ability to process unstructured and structured data with similar performance. Further details are provided in the Methods section and Supplementary Information.

# DNA dataset

Based our methodology, we partitioned the dataset into two parts. The first is a pilot dataset containing random information encoded into 1,000 sequences to analyze the noise characteristic, and the second is the main dataset of 110,000 encoded sequences, termed test dataset. The two datasets were synthesized by Twist Bioscience and obtained at different dates for both the synthesis and sequencing steps to make sure that they are as independent as possible.
The test dataset was composed of 110,000 encoded sequences containing data from several sources: image, audio, text, and random information which was done to examine our approach on several different data modalities. The test dataset was purposefully split into two files, each containing 55,000 encoded sequences. The first is a compressed (zipped) folder with three modalities: an image, a 24-second audio snippet, taken from Neil Armstrong's iconic moon landing, and a text file from the DNA Data Storage Alliance[33]. In total, the size of this compressed file is about 1.5MB. The second file contained about 1.5MB of random information bits. The data is shown in Fig. 2a-d. This partition was made to allow for the examination of random data in parallel to structured data.
Sequencing was done using two methods, Illumina miSeq and Oxford Nanopore MinION. This was done to examine our approach on different sequencing technologies each with its own pros and cons. Prior to each sequencing, we performed PCR amplification using the standard protocol of Q5 enzyme for 12 cycles[34]. Sequencing with Illumina miSeq produced 528,636 reads for the pilot dataset, with an error rate of 0.079% and an average cluster size of 529. The test dataset had 3,215,249 reads, 0.123% error rate, and an average cluster size of 29. The paired-end reads were merged using the PEAR software tool[35].
Sequencing with Oxford Nanopore MinION was done using the Ligation Sequencing Kit LSK 110 with flowcell 9.4.1[36]. The pilot dataset consists of 2,805,705 reads, with an error rate of 4.6% and an average cluster size of 754. The test dataset was sequenced twice, the first flowcell produced 4,341,575 reads, an error rate of 4.1%, and an average cluster size of 13. The second flowcell produced 3,065,455 reads, with an error rate of 5.07% and an average cluster size of 8. The combined two flowcells produced 7,407,030 reads, an error rate of 4.47% and an average cluster size of 21. Two sequencing runs were performed to ensure the average cluster size was sufficiently large.
The noise characteristics of both the pilot and test datasets for both sequencing methods were obtained using the SOLQC tool[32] and are shown in Fig. 2e-f where we see a good fit between them. As our approach is based on generating simulated data for training, this illustrates the premise that it is possible to overcome challenges such as domain transfer between the simulated data used for training and the actual data during operation based on our approach. Additional details on our datasets can be found in the Supplementary Information.

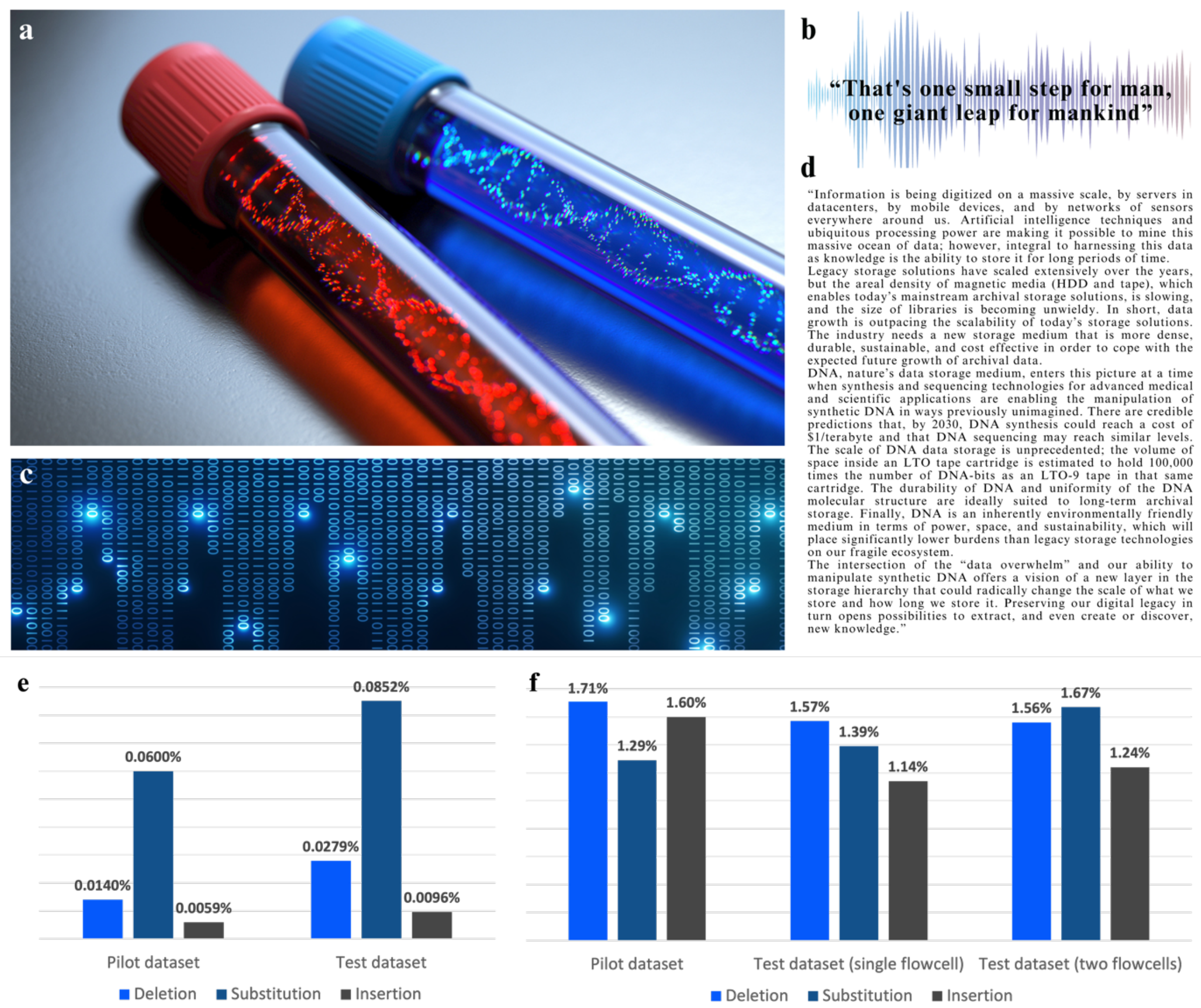

**Fig. 2 | Data used for DNA experiments. a-d** are an image, audio file, random bits and text accordingly. **e, f,** show statistical analysis of the errors found in the Illumina and Nanopore datasets accordingly. We deliberately chose these data modalities to examine the performance of our method under different conditions.

# Results

To examine and compare the accuracy and performance of the DNAformer, we used both of our datasets, Illumina and Nanopore (termed after the sequencing technology used), as well as 4 additional publicly available datasets. The previously published datasets differ in their synthesis and sequencing technologies, leading to different noise characteristics. To allow a fair comparison of the different reconstruction algorithms on the datasets, all of them were clustered using our binning approach. Since some of the datasets do not include indices in their encoded sequences, we used the first unique symbols of these sequences as they were indices in the clustering process.  The tested datasets were synthesized by Twist Bioscience, except the dataset by Grass et al.[12] which was synthesized by CustomArray. The sequencing technology for Grass et al.[12] and Erlich et al.[14] was Illumina miSeq, while Organick et al.[15] used Illumina NextSeq and Srinivasavaradhan et al.[7] used Oxford Nanopore MinION. Additional details on the publicly available datasets used can be found in the Supplementary Information.

For each dataset, we compared our method with 5 additional previously published algorithms. The first three algorithms use a symbol-wise majority vote approach in which the most frequent symbol in each position along the read is considered as the algorithm's prediction. These algorithms are based on the Bitwise Majority Alignment (BMA)[24] and are termed BMA Lookahead[37], Divider BMA[6], and VS algorithm[23]. The fourth method is the Iterative algorithm[6]. This algorithm uses dynamic programming methods to detect sequential patterns (small sub-sequences of the reads) that were observed in the reads and then uses the concatenation of the most frequent one as the algorithm's output. The fifth method is the hybrid algorithm[6], which uses the iterative algorithm and the Divider BMA algorithm, based on the given cluster and its estimated error probability.

Additionally, Srinivasavaradhan et al.[7] presented two trellis-based methods, the first is theoretical and the second is a heuristic improvement over the first, termed Trellis BMA. A comparison of the DNAformer with the Trellis BMA algorithm can be found in the Supplementary Information. A similar theoretical approach was also studied by Lenz et al.[8].

Notable work by Yazdi et al.[16] presented a coding scheme to correct errors caused by Nanopore sequencers and relied on coding constraints (such as exactly 50% of GC-content). As these constraints are not satisfied in our datasets, we did not include this method in our report.

A comparison of our approach to leading reported methods is provided in Fig. 3a, where the DNAformer achieves state-of-the-art (SOTA) results. On our Illumina and Nanopore datasets, the DNAformer achieves a failure rate of 0.0055% and 1.65% respectively. Additionally, our method achieves a failure rate of 0.02%, 0.66%, 0.17%, and 14.58% on the datasets provided by Erlich et al.[14] Grass et al.[12] Organick et al.[15], and Srinivasavaradhan et al.[7], respectively. We note that the closest method in reconstruction ability to the DNAformer is the iterative method[6].

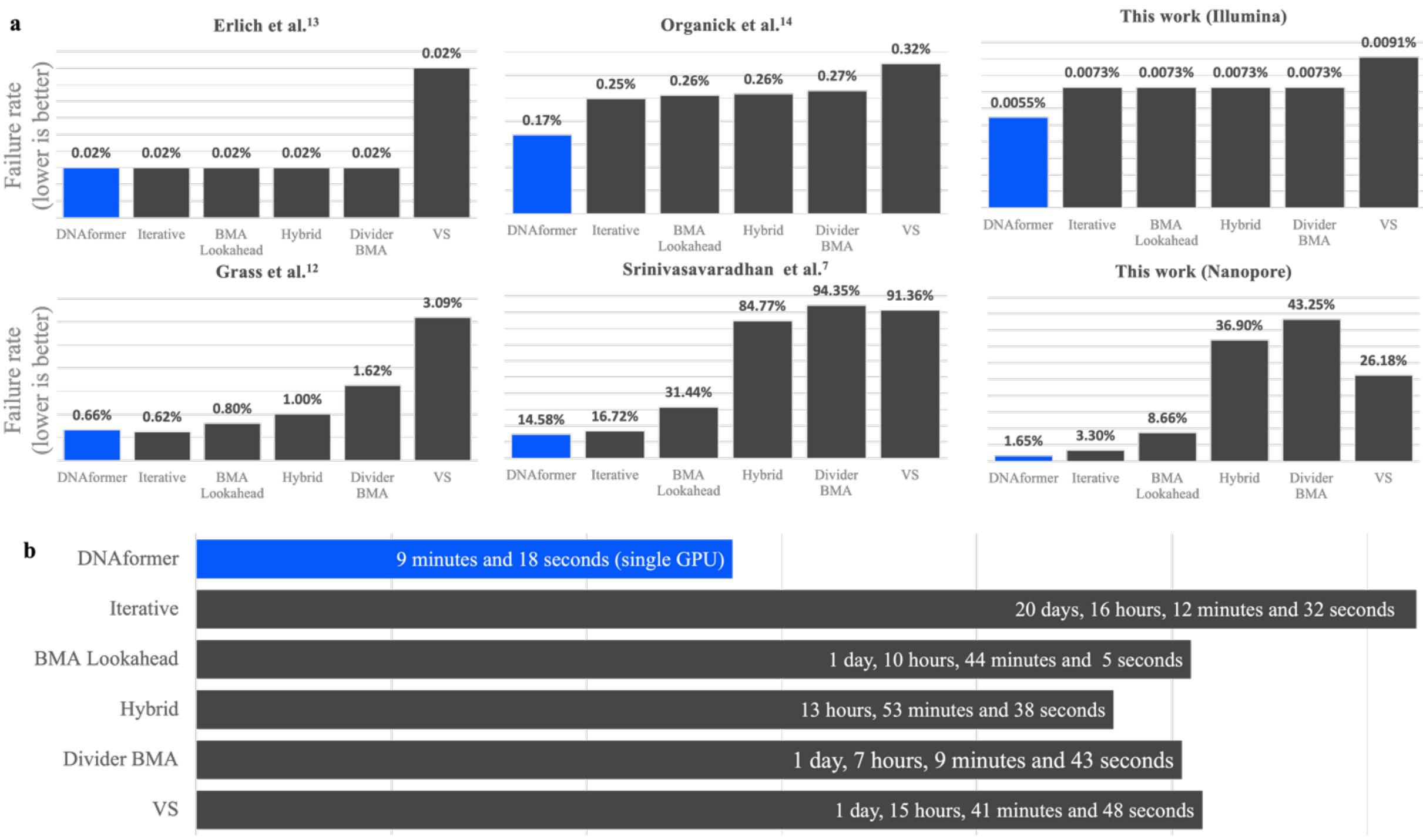

**Fig. 3 | Comparison of the DNAformer to SOTA DNA reconstruction methods. a,** shows the failure rate over several publicly available datasets as well as the Illumina and Nanopore test datasets, where the DNAformer achieves SOTA results. **b,** shows the estimated reconstruction for 100MB of information, where we see our method achieves significant reduction in the required time.

We further compare the accuracy of the DNAformer on different data modalities in our dataset. The first file includes text, an audio message, and a photo, while the second is a file consisting of random bits. Our results, provided in the Supplementary Information, show similar accuracy for both files. Thus, showing our method does not rely on the underlying semantics or structure in the data, but rather on the noise characteristics from the synthesis and sequencing processes. In addition to the reconstruction ability, it is also important to examine the inference speed of the different methods, as storage-based applications greatly favor fast processing speeds. The results are reported in Fig. 3b, where we compare the time, each method will take to reconstruct 100MB of data. Our method greatly improves on current methods by several orders of magnitude. When compared to the second-in-performance method (i.e. the Iterative method), we improve the speed by x3200. Meaning, the DNAformer simultaneously improves both reconstruction accuracy and speed without the natural tradeoff that usually exists between the two.

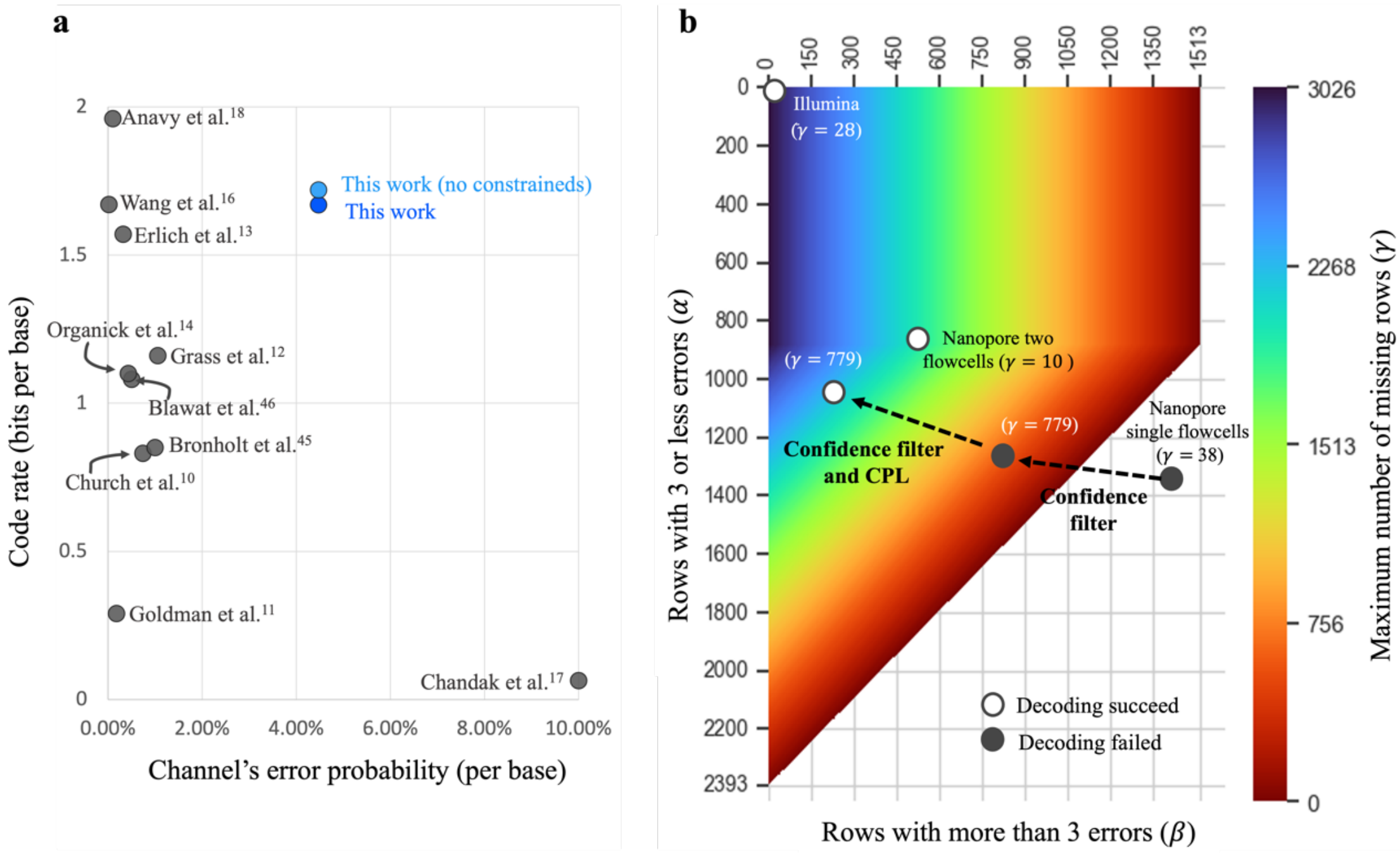

**Fig. 4 | Evalution of information retrival performance. a,** comparison of the code rate versus the channel's error probaility of different coding schemes that were used in DNA-based storage experiments. Our results show competetive code rate even while operating in higher noise regimes. **b,** heatmap of the errors regime where our coding scheme is able to operate susccessfuly. The specific instances of our three datasets are shown on top and illustrate a failure/success of the retrival as well as their associated safety margin.

To guarantee the retrieval of binary data, our method uses a TP based, modular coding scheme. When comparing different coding schemes, an important parameter to consider is the code rate which signifies how efficient the code is in terms of redundant symbols. Furthermore, the code rate needs to be evaluated alongside the channel's error probability as increase of this parameter makes the retrieval more challenging. Fig. 4a shows a comparison between different coding schemes where we see our work achieves a high code rate in addition to being utilized in a relatively high noise regime.

Apart from guaranteeing information retrieval the coding scheme presented in this work also introduces safety margins which signify how robust is the DNAformer under specific working conditions. As there are three main categories of errors from the DNAformer, where some are more difficult to correct than others, our code has been designed to support a range of values for each error type. Fig. 4b shows a heatmap with the three types of errors the code can correct and $\alpha, \beta, \gamma$ denote the quantities of each type of error. We see that when tested on the Illumina dataset, the DNAformer is well within the safety margins for information retrieval. When tested on the Nanopore two flowcells, the DNAformer can retrieve the information, however, the safety margin has been decreased. When tested on the Nanopore single flowcell, we see that the DNAformer is not able at first to retrieve the information. Following the method shown in Fig. 1, when applying both the confidence filter and the CPL, the DNAformer can successfully retrieve the information. Additional experimentation results, ablations studies, and comparisons are presented in the Supplementary Information.

# Discussion

When considering commercial, real-life applications of DNA-based storage, several system-related considerations arise, such as robustness, scalability, run-time, and compute requirements. As this work presents an end-to-end solution to the DNA information retrieval problem, several individual components in the pipeline have been re-designed. Additionally, focus was placed on the interaction between the different components which is critical to a successful operation. The results show SOTA performance in accuracy and run-time speed on several publicly available datasets in addition to new datasets provided in this work. Extensive experimentation and ablation studies are provided in the Supplementary Information showcasing the modular nature of this approach and how it can be modified to different applications and settings.

Our solution is based on the combination of TP-based ECC wrapping a transformer-based DNN while considering the system-related considerations mentioned earlier. To overcome the runtime limitations of current DNA-based storage pipelines, while using the inherent redundancy of the DNA-based storage channel, our design uses a naïve and very efficient method for clustering. However, this comes at the cost of noise within each cluster. Therefore, the information retrieval pipeline needs to be able to overcome this type of noise.

DNNs are a good fit for these requirements due to their parallel computational nature and Graphics Processing Unit (GPU) implementations, which allowed us to achieve inference time several orders of magnitude faster than previous solutions. However, the current cost of producing a large volume of real data for training such a model is high. In addition, since there is a need to employ an ECC prior to the DNA synthesis process, some changes in the design of the code will require to generate a new dataset. For these reasons, we turn to simulated data to train our model.

During training, we utilized 1.4B simulated DNA reads, dwarfing any existing datasets for DNA-based storage. The cost to create such a large dataset from real DNA (i.e., not simulated data) is estimated at over $10M, far greater than most labs' budget. Showing how our method paves the way for a realistic, large-scale application of DNA-based storage. Our training methodology uses a small amount of data from real experiments to model the errors, from which we can create an unlimited amount of simulated training data.

The proposed coding scheme introduces different attributes. First, we consider the entire information retrieval pipeline, which allows a simplification of the channel. Our approach strips away complexities involved with correcting insertions and deletions, making the process simpler. Second, by strategically leveraging a TP code, instead of the inner-outer code approach, we demonstrated that the integration of the clustering and reconstruction steps into the decoding process leads to a significant reduction in the required number of redundant symbols for error correction. Third, we offer a modular way to incorporate various constrained codes into our scheme, a known challenge in Coding Theory[13,38]. This allows for a versatile means of adapting to different conditions or requirements, enhancing the adaptability of our approach across various scenarios.

From a system-design perspective using the confidence filter and CPL algorithm increases the safety margin and allows our method to expand its ability to solve difficult cases. For example, higher noise regimes or smaller clusters sizes, a desired trait in terms of cost and time.
Our proposed method of integrating ECC and constrained codes extends outside the domain of DNA-based storage and can be useful in other channels as well. Moreover, the utilization of a TP based scheme that relies on similar ideas can be used for scenarios in which a DNN (or any other algorithm) can be applied to parts of the data pre-decoding.

# Conclusion

In this work, we present a scalable method for DNA-based storage that overcomes some of the major bottlenecks in current solutions for balancing failure rate and run-time. Our method combines a DNN and ECC to leverage the sequencing and synthesis inherent redundancy without compromising on performance. Furthermore, our solution reduces the required ECC redundancy and the required number of reads for error-free recovery of the information. Our results showed that DNNs can significantly improve the decoding process in a DNA-based storage system and shorten a DNA-based storage system response time by several orders of magnitude. From a broader perspective, our DNN-ECC approach overcomes a major bottleneck on the path to large-scale commercial applications of DNA-based storage.

# Methods

Our method is an end-to-end solution to the DNA information retrieval problem, as shown in Fig. 1a. As part of this approach, it was divided into several components, each with its own functionality: encoding of sequences prior to DNA synthesis, clustering of reads post sequencing, reconstruction and decoding back to the original data. In addition, the interplay and interface between the different components was also taken into consideration as part of a complete system-level, optimized solution. The solution combines coding theory and deep learning methods to create a holistic and coherent pipeline to encode and decode the DNA data. One of the merits of DNA-based storage is high data density, meaning a scalable storage system needs to be able to quickly process arbitrary large files. To create a scalable method, we do not require the entire file to be processed simultaneously and design our method to process data in smaller batches, hence, our solution can be adapted to random access purposes. Our suggested solution encodes a block of 1.53962MB into 55k designed encoded sequences of length 140. Each strand is composed of 12 bases for index, including a single base that serves as a file identifier. The remainder 128 bases were allocated for the binary data and the redundancy from the ECC and constrained code. This length was chosen to allow for efficient DNN processing on a GPU during the reconstruction step and the coding scheme during the encoding and decoding steps.

## Coding scheme

Our coding scheme is a modular pipeline composed of several components, each addressing a different purpose: index encoding, diagonal column encoding, constrained code and TP code. The complete encoding and decoding pipeline is designed to integrate these four components in an interleaving order that allows each of them to achieve its designed goal without hindering the others.

There are three main considerations in the design of our coding scheme. First, our solution assumes that the decoding is performed after the clustering and reconstruction algorithms. By design, these steps eliminate most of the deletions and insertions errors, and the output has the same length as the encoded sequences. Hence, after this step, we need to only take care of substitution errors and missing predictions due to either missing clusters or our confidence filter, which is easier to solve in terms of coding theory and requires less redundancy.

As seen in Fig. 3a, when the clustering and reconstruction perform well, a high number of error-free predictions can be passed to the decoder, so the decoder only needs to correct errors in a small fraction of the predicted sequences, rather than in all of them. Therefore, a main component of our code is a TP based coding scheme that can correct up to a specified number of errors in up to a specified fraction of the predictions, allowing for a significant reduction in the redundancy needed to ensure error-free reconstruction of the data.

To maintain error-free retrieval of the information, our code also uses a diagonal column encoding which serves as an outer code that can correct the remaining erroneous predictions and overcome the missing ones. This code is implemented using an RS code which is applied on the encoded sequences that store the information. As some sequencing and synthesis technologies are more error-prone at the beginning and end of each read[32,39] the RS code was designed diagonally. Meaning, the information at the beginning and end of each read will be encoded together with the information in the middle to create a more uniform distribution of the errors.

The second consideration in our design is that our clustering step is essentially a binning algorithm that uses the index part of the reads and matches them to the valid indices that were used in the designed sequences. Hence having errors in the index can cause a read to be ignored or misclassified. As misclassified reads are more problematic during the reconstruction step, we use an index encoding that is based on a pre-calculated set of indices that are of edit distance 3 or more from each other.

Our third consideration relates to constrained coding for DNA-based storage. The different sequencing and synthesis technologies lead to different constraints that should be addressed, depending on the specific technology. For example, in our datasets we selected the mapping to avoid the long homopolymer (consecutive repetition of the same base) of length 5 or more and GC content of each encoded sequence between 45% to 55% with high probability. These constraints were selected to preserve the stability of the synthesized strands and mitigate their error rates, considering the technologies that are used in our experiments[40–42].

Additionally, recent work demonstrated that in some cases, it is better to not impose any constraint[41]. Hence, to keep our code flexible for different use cases we designed a block-based constrained code in a way that is almost independent of other components of our scheme. This allows an easy adaptation to other constraints or a removal of the constrained code component entirely. In our design, removing the constrained code improves the information rate by 7.6%.

## Clustering

From a system-level optimization perspective, this step was optimized for speed rather than accuracy. We adopt a naïve clustering approach based on simple and fast binning of the reads based on their index. The goal in our design for this step is to overcome slow clustering methods[21,22] and reduce the processing time. However, this comes at the cost of noise, introduced by errors in the index of each read, which causes some of the reads to be wrongly clustered (i.e. false reads) or to be dropped during the clustering step. On a system-level basis, we overcome this noise in the reconstruction step using a DNN.

## Reconstruction

Designing a method which combines a DNN and ECC requires the ability to iterate between the two parts during the design phase. That is, the coding scheme and the DNN are coupled together to guarantee a specific set of success metrics. However, creating a different training dataset for each coding scheme modification is a costly and resource intensive process. For example, using previous DNA-based storage systems[14,43], the estimated cost of synthesizing 1GB of data is roughly $3-5M.

Due to this fact, we turn to simulated data for training our DNN. The main challenge when using simulated data for training is the generalization to real-world settings after the model is trained. To overcome this issue, we construct a data generator based on statistics from real-world experiments[32,44]. These statistics contain the error probabilities which are used to generate the reads of each label.

### Simulated data generation

Our reconstruction approach uses a DNN which predicts a single label sequence from a cluster of reads. Since the data is randomly sampled, each cluster can vary in size, and each read can vary in length due to synthesis and sequencing errors. Moreover, some clusters can suffer from higher error rates compared to others. Our method utilized simulated data only during training, meaning, we did not use real data during the training of the models. Pseudo code for each iteration of the training process using our simulated data generator is given below:

1. Draw a random sequence of letters based on the 4-ary {ACGT}.

2. Encode the sequence using some error-correcting and constrained coding scheme (optional).
3. Draw a random number of reads and a random number of false reads.
4. For each read:
    a. Draw random deviation from the modeled error probabilities.
    b. Inject simulated errors for each read based of the error model[45].
5. Batch several clusters.
6. Forward-backward pass through the DNN.

## Data preprocessing

The DNAformer architecture processes the reconstruction of a cluster of reads. Meaning, a set of reads is processed simultaneously to reconstruct the suitable encoded sequence. Preparation of the reads includes filtering short and long reads beyond a specified design parameter. This ensures that highly corrupted reads will not affect the reconstruction process. Following, a simple one-hot encoding and padding are used to prepare the data to the model's encoder and make sure all reads are of the same length.

## Model architecture

Our model uses a combination of convolutions and transformers. We adopt the concept of early convolutions before a transformer block to improve training stability and performace[46]. The model is structured as a Siamese network where the two branches share weights and are different by reversing the symbol order prior to the branch termed 'right branch' in Fig 1b. This is done since the padding used in the preprocessing step is concatenated only at the end of each read. An alternative approach will be to align all the copies to one another using sequence alignment methods[47]. However, these methods come at the cost of 'in-series' computation and processing time and therefore we decided to avoid them.
Instead, we created an alignment module whose purpose is to learn the required alignment for each read independently. The alignment module uses an Xception[48] inspired architecture with depthwise separable convolutions and multiple kernel heads. The purpose of using multiple kernels in the embedding layer is to allow the model to learn different shifts caused by deletion or insertion errors. Following this module, we sum over the cluster dimension with the goal of improving the model's robustness to differences in cluster size. This operator has the effect of Non-Coherent Integration (NCI) and aims to increase the Signal to Noise Ratio (SNR) of the data.
After the NCI aligner layer, the model includes an embedding module whose architecture is similar to the alignment module, however, now the operations are employed on the whole cluster and not on each read independently. The goal is to learn correlations between the different reads and prepare the data for the Transformer module. In addition, the embedding module outputs a sequence with the required output length and larger feature space.
The transformer module is a multi-head transformer architecture, used with Multi-Layer Perceptron as feedforward layers[49]. We do not use position embedding in this module. After the last transformer block, a linear module is used to reduce the number of features to 4 which represent one-hot encoding for the DNA representation.
The fusion layer is a vector of learnable parameters with a length of the required encoded sequence. This layer combines between the predictions of the two branches into a single prediction prior to a softmax operator which transforms this representation to probabilities. Additional ablation studies and details on the architecture are provided in the Supplementary Information.

## Loss function

To train our model a combination of cross entropy and consistency loss was used and are shown in Eq. (1), (2), and (3):

$$\mathcal{L} = \lambda_{\mathrm{ce}}\mathcal{L}_{\mathrm{ce}} + \lambda_{\mathrm{consistency}}\mathcal{L}_{\mathrm{consistency}} \quad (1)$$

$$\mathcal{L}_{\mathrm{ce}} = -\frac{1}{\mathrm{n}}\sum_{\mathrm{n}} y_{\mathrm{n}}\log(\theta_{\mathrm{n}}) \quad (2)$$

$$\mathcal{L}_{\mathrm{consistency}} = 0.5 \cdot \left(\mathcal{L}_{\mathrm{ce}}(\theta_{\mathrm{Left\ Branch}}) + \mathcal{L}_{\mathrm{ce}}(\theta_{\mathrm{Right\ Branch}})\right) \quad (3)$$

Where $\lambda_{\mathrm{i}}$ are hyperparameters, $\theta_{\mathrm{i}}$ are the model prediction probabilities and $y_{\mathrm{n}}$ are the labels. Left and right branches refer to the Siamese architecture. The formulation for consistency loss which yielded best results is shown in Eq. (3) and aims to minimize the average prediction of the two branches, giving the model the freedom to adapt to changing noise characteristics along the sequence length.

## Training details

Data generation and training was implemented in Pytorch, optimizer used was Adam with $\beta_1 = 0.9, \beta_2 = 0.999$, batch size 64 and learning rate utilized cosine decay from $3.141 \cdot 10^{-5}$ to $3.141 \cdot 10^{-7}$. A single A40 GPU was used during training and inference. Training took 50 epochs for the Illumina experiment and 180 epochs for the Nanopore experiment, each containing 1M clusters and an average number of 8 DNA reads per cluster.

## CPL

The CPL algorithm is a second reconstruction step used on clusters with low confidence from the DNN output. The algorithm receives a cluster of reads and its goal is to predict their encoded sequence. Since the clusters that are sent to the CPL algorithm have low confidence, often the DNN predictions of these clusters are not correct. Therefore, we built the CPL to not use either the inference created by the DNN nor the prior knowledge of the sequencing and synthesis error rates. The algorithm works directly on the clusters as they were obtained by the clustering algorithm.

First, the CPL algorithm takes the first read from the cluster, and then uses dynamic programming to calculate the edit distance of the first read from any of the other reads in the cluster. This step is used to estimate the errors that occurred in the reads of the cluster. Based on this calculation, the algorithm creates vectors of edit operations (deletions, insertions, and substitutions) that describe how to transform the first read to any of the other reads in the cluster using edit operations. These edit-operations vectors represent estimations of the number of occurrences of each edit error in the cluster and their location with respect to the first read. Next, the algorithm creates a directed acyclic graph based on these estimations of the errors. The vertices in the graph are the symbols of the first read, or the error events that were predicted in it based on the calculations. The edges in the graph connect any two vertices that correspond to symbols/errors which occur in adjacent locations. The weights of the edges are defined based on the occurrences of the two connected vertices in the edit-operations vectors. Finally, the algorithm returns the longest path, that represents the sequence with maximum probability based on the algorithm's estimations.

## Confidence filter and safety margin

One of the key features in our suggested solution is the confidence filter, which is a function that allows us to decide whether to rely on the DNN's output or not. The confidence filter is used for two purposes, the first is classifying highly erroneous DNN predictions as erasures, where in our pipeline, this equates to a missing cluster. This is beneficial as correcting erasures requires less redundancy than correcting substitutions of entire sequences. The second purpose is to decide which of the clusters should be sent to the CPL algorithm.

The main component of the confidence filter is a confidence function which operates on the soft output of the DNN. The soft output of the DNN is a $4 \times L$ matrix $M$, in which the 4 entries of the i-th column can be thought as the probabilities that the $i$-th symbol of the prediction is the corresponding symbol ACGT and $L$ is the encoded sequence length.

In the confidence function we utilize this property, specifically, we use the arithmetic mean of the maximal value in each column, which represents the probability that the predicted symbol is correct. This mean value is defined as $m(M) \triangleq \frac{1}{L}\sum_{j=1}^{L} \max\{M_{1,j}, M_{2,j}, M_{3,j}, M_{4,j}\}$, where $M_{i,j}$ is the value of the $i$-th entry in the $j$-th column of $M$. Intuitively, smaller values of $m(M)$ correspond to cases in which the DNN is less confident in its output.

An additional property that is considered in the confidence filter is the cluster size. More precisely, our confidence filter considers the fact that the output of the confidence function of smaller clusters, i.e. lower SNR, tends to be lower in general and integrates this property to the final filtering decision. The confidence function is defined as $c(M, cluster\ size) \triangleq m(M)^{2 \cdot cluster\ size}$. The confidence filter evaluates the following conditions:

1. If $c(M, cluster\ size) \leq confidence_{threshold}$ and $cluster\ size \leq cluster\ size_{threshold}$ the cluster is classified as an erasure (i.e. missing cluster).
2. Otherwise, if the CPL is incorporated in the pipeline, $c(M, cluster\ size) \leq confidence_{threshold}$ and $cluster\ size > cluster\ size_{threshold}$ then the cluster is passed to the CPL algorithm.
3. Otherwise, the confidence filter trusts the DNN output and passes it as it is to the decoder.

The optimization for the accuracy vs. runtime tradeoff is controlled by the values of $confidence_{threshold}$ and $cluster\ size_{threshold}$ and was performed using the pilot datasets. The performance of the DNN on the Illumina dataset was very good and did not require any other mechanism, therefor we set $confidence_{threshold} = 0$ and $cluster\ size_{threshold} = 3$. The performance of the DNN on the Nanopore single flowcell did require additional mechanisms to ensure successful information retrieval and robust safety margin. Hence, we chose a safety margin of 1%, which yielded the values of $confidence_{threshold} = 0.7$ and $cluster\ size_{threshold} = 4$. Evaluation was performed on the test datasets with the details provided in the Supplementary Information.

# Supplementary Information for Deep DNA Storage: Scalable and Robust DNA-based Storage via Coding Theory and Deep Learning

Daniella Bar-Lev*, Itai Orr*, Omer Sabary*, Tuvi Etzion and Eitan Yaakobi
Department of Computer Science, Technion – Israel Institute of Technology, Haifa, Israel
*These authors contributed equally

# Supplementary results

## Effects of the data structure on the error rate

To test the robustness of our suggested solution, we compared the failure rates of each of the files separately. The results are summarized in Supplementary Table 1 where we see that the DNAformer shows similar accuracy on both files. The failure rate measured in all three datasets is similar between the two files suggesting that the success of the retrieval pipeline of DNAformer does not depend on the entropy of the stored information**.**

| File | Test dataset Illumina | Test dataset Nanopore \| single flowcell | Test dataset Nanopore \| two flowcells |
|---|---|---|---|
| File 1 - photo, audio, and text | 0.056% | 3.8% | 1.61% |
| File 2 - random bits | 0.056% | 3.82% | 1.49% |

**Supplementary Table 1 | Data modality effect on the failure rate.** The results show our method is invariant to the data modality, implying that the DNAformer does not use the underlying structure in the data. Rather, we learn the noise characteristics of the channel and adapt to each type of synthesis-sequencing pair.

## Effects of cluster size on the error rate

Evaluation of the reconstruction accuracy of the DNAformer as a function of the cluster size is provided in Supplementary Fig. 1. The results are shown on the test dataset, considering the read obtained by Nanopore MinION from two flowcells. For any cluster size $3 \leq t \leq 16$, we sample $t$ copies from the obtained clusters that have at least $t$ copies. The results of the reconstruction accuracy are shown by the failure rate, the hamming error rate and edit error rate.

It can be seen that larger clusters correlate with higher accuracy in all of the tested algorithms. However, the improvement decays for cluster of size more than 15. For example, when comparing the success rate of clusters of size 15 with the success rate of clusters of size 16 the improvement of all the tested algorithm was less than 0.8%. Moreover, the improvement of the success rate of the DNAformer is less 0.1%. Additionally, since the DNAformer was implemented on GPU, for inference efficiency purposes, the maximal cluster size was selected to be 16. It can be further seen that for clusters of size 4 and below, the results of most of the algorithms are relatively poor, up to 50% success rate, average edit distance of 1 to 9, and Hamming distance of at least 5. Thus, the cluster size threshold of the confidence filter was selected to be 4. Further details on the design of the confidence filter are provided in the appropriate subsection.

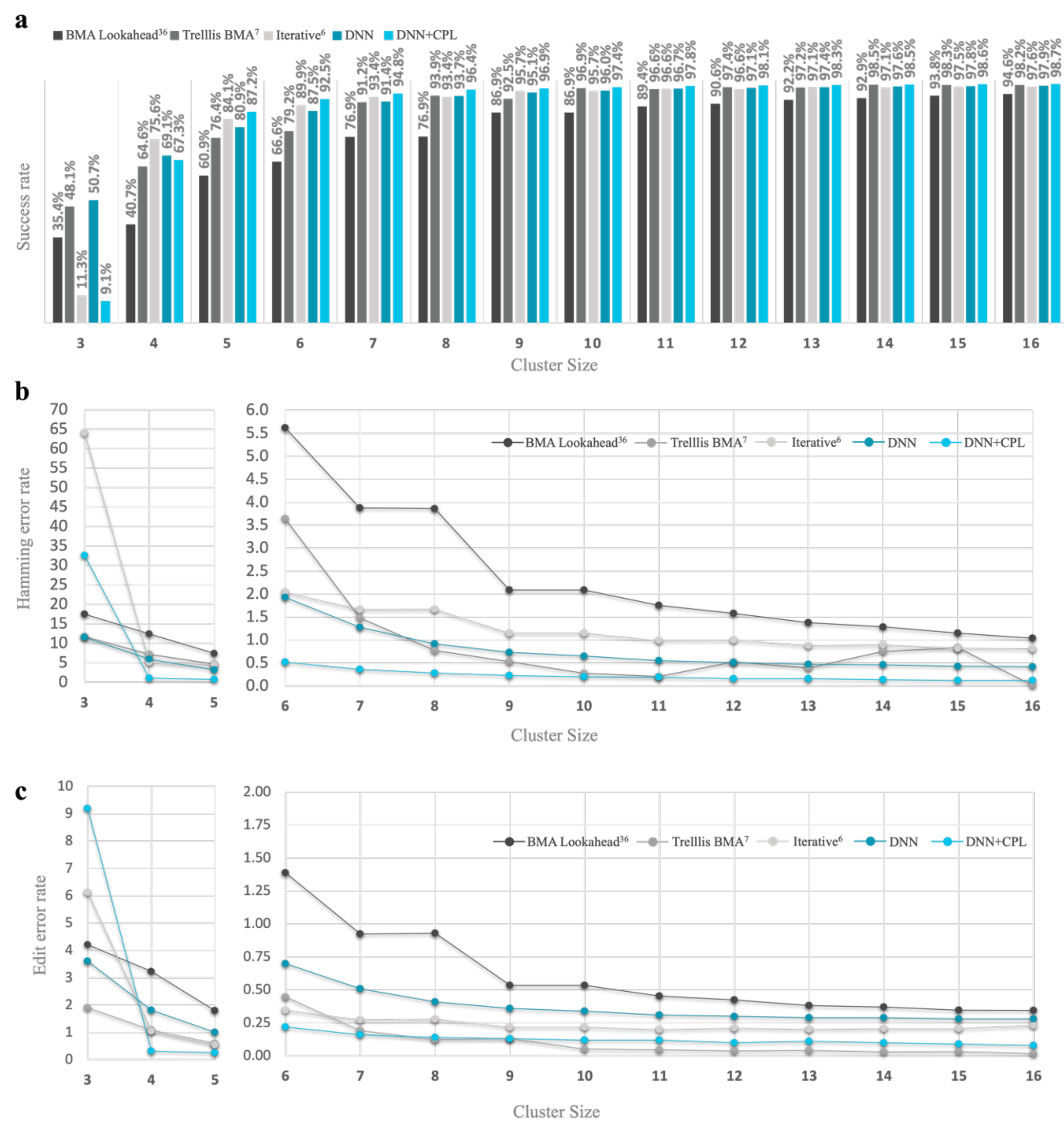

**Supplementary Fig. 1 | Cluster size effect on the error rate of the DNAformer. a,** success rate as a function of the cluster size. We see the results improve as the cluster size increases, an attribute associated with the cluster's SNR. **b, c,** Hamming and edit error rates as a function of the cluster size accordingly. The results show that larger clusters enhance the reconstruction accuracy, however the enhancement decays when the cluster size reaches 16 and thus this size was selected as the maximal cluster size for the DNAformer. Furthermore, it can be seen that the accuracy is lower for clusters of size 4 and below, and thus the size 4 was selected as the threshold for the confidence filter.

## Analysis of the clustering step

Supplementary Table 2 presents the results of the binning algorithm on our datasets. It can be seen that the binning algorithm clustered between 99%-100% of the reads that were obtained by Illumina sequencing, while only 26%-33% of the Nanopore reads were clustered successfully.

| Dataset | | Number of reads | Number of clustered reads | Number of clusters |
|---|---|---|---|---|
| Pilot | Illumina | 528,636 | 528,636 (100%) | 1,000 |
| | Nanopore | 2,805,705 | 753,888 (26.86%) | 1,000 |
| Test | Illumina | 3,215,249 | 3,183,840 (99%) | 109,944 |
| | Nanopore first flowcell | 4,341,575 | 1,446,602 (33.31%) | 109,928 |
| | Nanopore second flowcell | 3,065,455 | 907,982 (30%) | 109,753 |
| | Nanopore two flowcells | 7,407,030 | 2,354,584 (31.78%) | 109,976 |

**Supplementary Table 2 | Results of the binning algorithm on the different datasets**. The right column is the number of non-empty clusters that were obtained by the algorithm out of 1,000 for the pilot datasets and 110,000 for the test datasets. In the second to right column, we see a high clustering percentage for the Illumina dataset while the Nanopore dataset exhibits lower clustering percentage due to its higher noise regime.

# Analysis of the accuracy throughout the retrieval pipeline

Supplementary Table 3 shows a detailed description of the reconstruction accuracy of each of the steps involved in the retrieval pipeline of the DNAformer as described in Fig. 1. It can be seen that the total success rates of the DNAformer were between 96.12% and 99.94%. Furthermore, for the Nanopore reads, when using a single flowcell the number of wrong predictions was 5,550 (5.05% of the clusters) and when using two flowcells the number of wrong predictions was 2,720 (2.47% of the clusters). Our confidence filter removed 2.9% of the clusters (single flowcell) and 1.1% of the clusters (two flowcells), while the CPL was performed on 52.4% of the filtered clusters (single flowcell) and 91.75% of the filtered clusters (two flowcells).

| Step | Test dataset \| Illumina | Test dataset \| Nanopore two flowcells | Test dataset \| Nanopore single flowcell |
|---|---|---|---|
| Number of tested clusters | 109,944 | 109,976 | 109,928 |
| Missing clusters | 56 | 24 | 72 |
| Wrong predictions of the DNN | 6 | 2,720 | 5,550 |
| Number of clusters filtered by the confidence filter | 0 | 1,189 | 3,197 |
| Number of clusters sent to the CPL algorithm | 0 | 1,091 | 1,676 |
| Number of clusters corrected by the CPL algorithm | 0 | 640 | 1003 |
| Number of clusters that were classified as erasures | 0 | 98 | 1,521 |
| Success rate from existing clusters | 99.99% | 98.34% | 96.18% |
| Total success rate | 99.94% | 98.32% | 96.12% |

**Supplementary Table 3 | Analysis of the accuracy throughout the retrieval pipeline.** The table shows the reconstruction results in each of the different components of our reconstruction pipeline. The results show that for the Illumina test dataset and Nanopore test dataset two flowcells, the DNN is able to complete the reconstruction on its own. However, for the Nanopore test dataset single flowcell the DNN alone does not guarantee information retrieval. By using the confidence filter and CPL our method is able to cope with this SNR regime and guarantee successful retrieval.

## Analysis of coding schemes

Supplementary Table 4 shows a comparison of the coding schemes and design parameters that were used in previous DNA-based storage experiments. The error rates presented in the table are either based on our analysis if the datasets are publicly available, or on the reported error rates by the authors if the datasets are not publicly available. Note that since the calculation of the error rates depends on the exact clustering method that is used, it is possible to get a slightly different values for these rates. The results show that even though our channel error rates are an order of magnitude higher compared to the previous works that used Illumina sequencing, our work still achieves amongst the highest information rates.

The work by Yazdi et al.[16] was the first to use Nanopore sequencing for DNA-based storage. Although the technology was more error-prone than it is today, they were able to recover the stored information. We did not include this work in Fig. 3 and Fig. 4 as the length of their encoded sequences was considerably longer and the number of clusters was considerable smaller than other works, as seen in Supplementary Table 4.

| Dataset | Data Size | Synthesis technology | Sequencing Technology | Encoded Sequences Length | Information rate (excluding primers) | Coding Technique | Channel error rate |
|---|---|---|---|---|---|---|---|
| Church et al.[10] (2012) | 0.65 MB | Agilent | Illumina HiSeq | 159 | 0.83 | Constrained Coding | 0.74% |
| Goldman et al.[11] (2013) | 0.63 MB | Agilent | Illumina HiSeq | 117 | 0.29 | Constrained Coding | 0.1774% |
| Grass et al.[12] (2015) | 0.08 MB | Custom Array | Illumina MiSeq | 158 | 1.16 | Constrained Codes + Inner-Outer Reed Solomon | 1.06%* |
| Bronholt et al.[50] (2016) | 0.15 MB | Not reported | Illumina MiSeq | 120 | 0.85 | Hoffman code+XOR (outer) | ~1% |
| Erlich and Zielinski[14] (2017) | 2.11 MB | Twist Bioscience | Illumina MiSeq | 152 | 1.57 | DNA fountain (Luby transform) + Reed Solomon | 0.32%* |
| Yazdi et al.[16] (2017) | 3 KB | IDT | Nanopore MinION | 1000 | 1.74 | Constrained codes + Deletion + Multiple sequence alignment | ~10% - 20% |
| Blawat et al.[51] (2016) | 22 MB | Agilent | Illumina MiSeq | 230 | 1.08 | Run length limited + Inner code CRC + Outer code RS | <0.5% |
| Organick et al.[15] (2018) | 200 MB | Twist Bioscience | Illumina NextSeq (Datasets) + Nanopore | 150 | 1.1 | Inner + Outer Reed Solomon | 0.43%* |
| Chandak et al.[18] (2020) | 11 KB | Custom Array | Nanopore | 165 | 0.063 | Inner CRC/Convolutional code + Outer Reed Solomon Code | ~10% |
| Wang et al.[17] (2019) | 379.1 MB | Twist Bioscience | Illumina HiSeq | 190 | 1.67 | CRC (single primer) + Repeat Accumulated Code | ~0.02% |
| Anavy et al.[19] (2019) | 6.4 MB | Twist Bioscience (Composite) | Illumina MiSeq | 194 | 1.96 | Reed Solomon (inner) + Fountain Code + Composite | <0.1% |
| This Work | 3.1 MB | Twist Bioscience | Illumina MiSeq + Nanopore MinION | 140 | 1.6 / 1.72 (no constraints) | Tensor Product Code + Flexible Block-based Constrained Code | 4.47%* |

**Supplementary Table 4 | A comparison of coding methods, synthesis and sequencing technology, and design parameters of previous DNA storage experiments.** We report that our work shows competitive results for the code rate while operating under higher channel error rate regimes than prior works. *The reported channel error rate is based on SOLQC[32]

### Comparison with the Trellis BMA algorithm

Supplementary Table 5 shows a comparison of our method to the Trellis BMA algorithm[7] and was performed separately due to its long running time. Therefore, we randomly selected 10,000 clusters from our Nanopore test datasets, each containing 16 reads. The running time of the trellis BMA algorithm on a 32 cores CPU was 61.2 hours, while the DNAformer runtime was 1.52 seconds on a single A40 GPU.
It can be seen that the DNAformer outperforms the Trellis BMA algorithm in all of the tested parameters. More specifically, the DNAformer improves the failure rate by over 40%, while the edit error rate is improved by 3.5%, and the Hamming error rate is improved by over 50%.

| | Trellis BMA | DNAformer |
|---|---|---|
| Failure rate | 3.3% | 1.97% |
| Edit error rate | 0.2266% | 0.219% |
| Hamming error rate | 0.677% | 0.313% |

**Supplementary Table 5 | Comparison of the reconstruction accuracy of Trellis BMA algorithm and the DNAformer**. The table shows that the DNAformer improves the failure rate, the edit error rate and the Hamming error rate of the Trellis BMA algorithm.

## DNA dataset

### Publicly available datasets

In this work we compared the performance using several publicly available datasets. The number of encoded sequences in the dataset by Grass et al.[12] was 4,991, while the pseudo-clustering algorithm obtained 4,982 clusters. The total error rate of this dataset was 1.06%, and the average cluster size was 528. The dataset by Erlich et al.[14] includes 72,000 clusters with a total error rate of 0.32% and an average cluster size of 178. The dataset by Organick et al.[15] contains 607,150 encoded sequences, their reads were clustered into 596,244 clusters with a total error rate of 0.43%. The average cluster size of this dataset was 32. Lastly, the dataset by Srinivasavaradhan et al.[7] consists of 9,954 clusters (created from reads of 10,000 encoded sequences) with a total error rate of 4.72% and an average cluster size of 17. It should be noted that in our reconstruction accuracy evaluations, the DNAformer used only up to 16 reads per cluster, while other algorithms used various larger cluster sizes. However, for runtime performance evaluations, we only compared clusters of size up to 16.

A detailed description of the error rate of the four different datasets can be found in Supplementary Fig. 2. The figure presents the substitution, insertion and deletion rates of each of our tested datasets. It can be seen that in the data sets of Erlich et al.[14], Grass et al.[12], and Organick et al.[15], the most dominant errors were deletion and substitution, while in the data set by Srinivasavaradhan et al.[7], the most dominant errors were insertion and deletion.

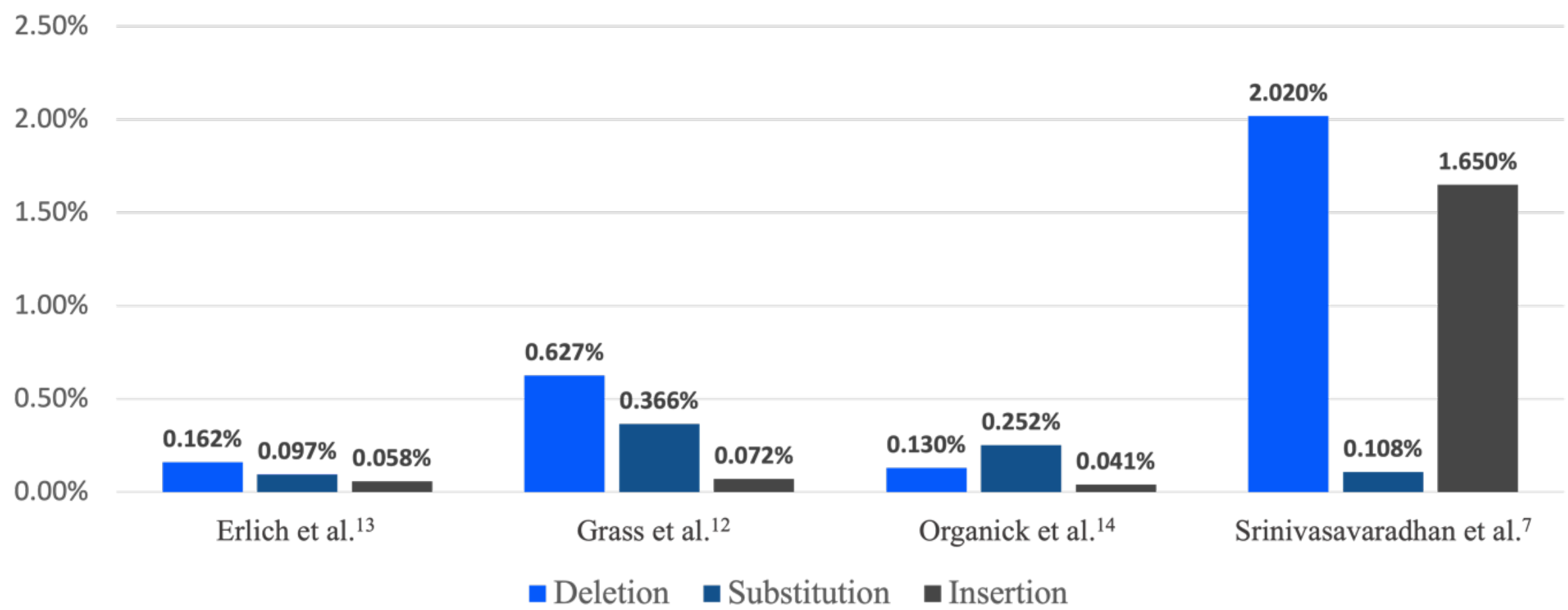

**Supplementary Fig. 2 | Error rates of publicly available datasets.** The figure shows the substitution, insertion and deletion rates of each of the tested data sets.

It should be also noted that there are three additional publicly available datasets that we do not include in our comparison. The work of Anavy et al.[19] presented a new approach for synthesis, in which the alphabet can be abstracted to include more than four symbols. This is done by utilizing composite letters that combine more than a single nucleotide on the same position, which make their data less relevant for our pipeline. The work by Yazdi et al.[16] consists of 17 encoded sequences, which is not enough for adequate error characterization by SOLQC[32], which is needed to train our DNN. Lastly, Chandak et al.[18] published a dataset that combines several different experiments together. As a result, their dataset contains multiple encoded sequences with the same index. This fact makes this dataset not suitable for our binning algorithm.

## Our Illumina and Nanopore datasets

Supplementary Fig. 3 shows the cluster size histogram for Illumina and Nanopore reads. The clusters were obtained using the binning step described in the Method section. All histograms are of bell-shape. For the Illumina datasets, shown in Supplementary Fig. 3a and Supplementary Fig. 3b the bell-center of the histogram of the pilot is around 550, and around 40 for the test dataset. For the Nanopore datasets, the bell center of the pilot dataset, shown in Supplementary Fig. 3c, is centered around 750. The Nanopore test datasets, for the case of a single flowcells and for the case of two flowcells, are shown in Supplementary Fig. 3d and Supplementary Fig. 3e and have the bell center around 12 and 21 accordingly.

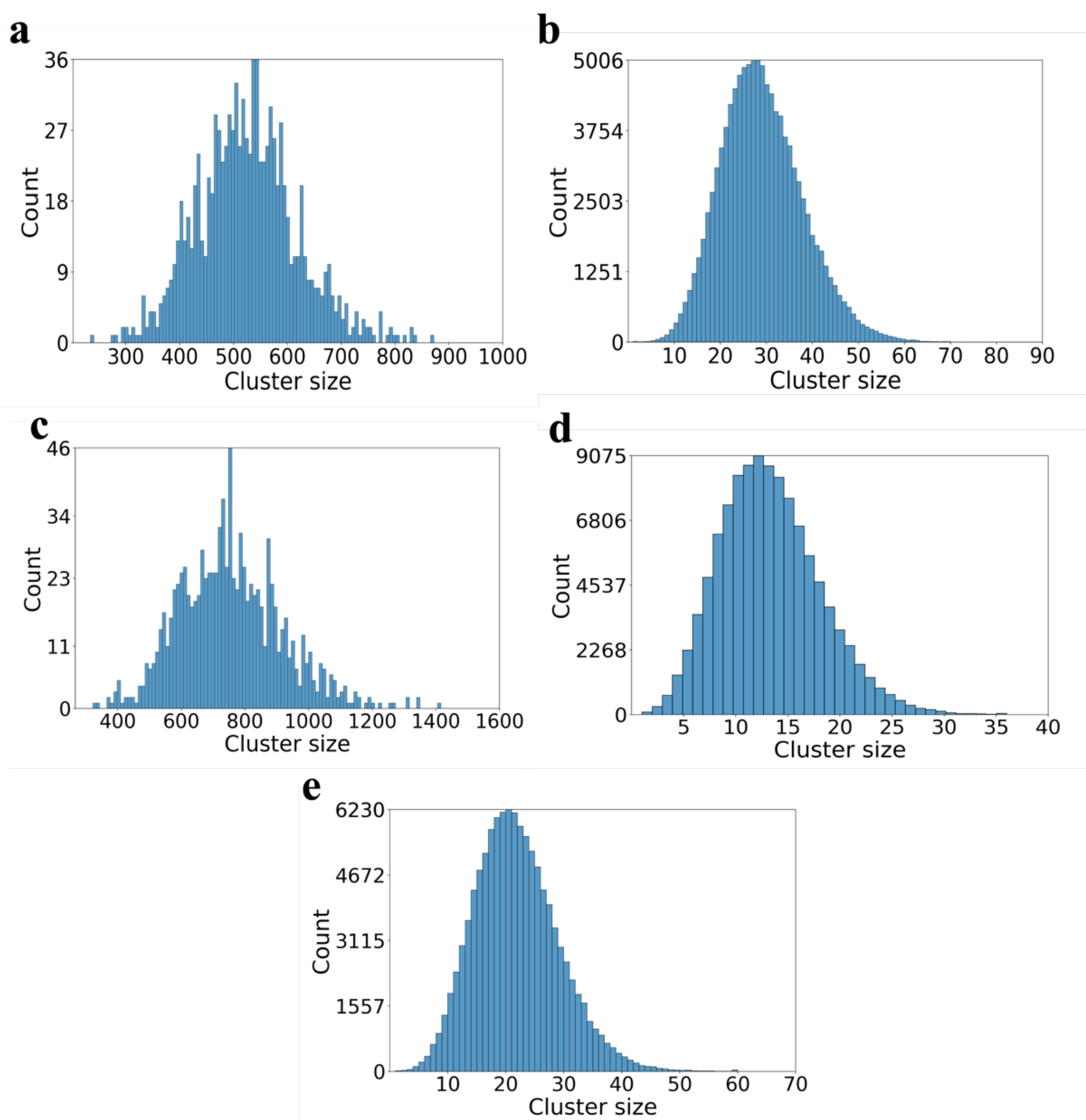

**Supplementary Figure 3 | Cluster size histograms of our data from Illumina and Nanopore datasets.** For all plots, the X-axis shows the cluster size and Y-axis shows the number of clusters that were obtained at this size. **a,** Illumina pilot dataset. **b,** Illumina test dataset. **c,** Nanopore pilot dataset. **d,** Nanopore test dataset single flowcell. **e,** Nanopore test dataset two flowcells.

# Clustering

Following the sequencing process, the first step of the information retrieval process is clustering of the obtained reads. In this step, the obtained reads are partitioned into small groups known as clusters based on their origin. The process is known to be computationally difficult, and since the number of obtained reads is high, performing the clustering may require long running time, even when using efficient methods[21,22].

The hash-based clustering method[22] takes a random short sequence of DNA, and partitions the reads based on the location of this randomly selected sequence. The clover clustering method[21] is based on a tree graph that is created from the reads. The edges in the tree estimate the distance between the reads and are used for clustering. Lastly, we also show the results of the perfect clustering method. In this method, we assume prior knowledge of the original encoded sequences and each read is mapped into the cluster of its nearest encoded sequence. For our comparison, we used a brute force algorithm to create the clusters.

In our suggested method, we utilize the defined indices in each read to enable a faster and more efficient clustering process. Essentially, we perform simple binning based on each read's index, a fast and efficient process. The binning algorithm examines the 12 first bases (the index) of each read and bins it into clusters based on the inspected index. If the 12 first bases of a read do not match to any of the possible indices, then the read is ignored. We compare our clustering method with previously published clustering algorithms and show that our retrieval pipeline is independent of our clustering method. That being said, using our binning method dramatically reduces the clustering time, while preserving competitive reconstruction accuracy of the DNAformer.

Examination of different clustering methods is presented in Supplementary Table 6. The evaluation was performed on a server with Intel(R) Xeon(R) CPU E5-2630 v3 2.40GHz. The results show the reconstruction failure rate and the runtime. It can be seen from the table that the running time of our binning algorithm is significantly smaller across all experiments.

It should be noted that the binning method considers the indices to obtain its clusters, while the perfect clustering method considers the encoded sequences to obtain its clusters. Therefore, when using these two methods, it is easy to match between an obtained cluster and an encoded sequence. However, when using the Hash-based clustering[22] and Clover clustering[21], a reconstruction algorithm should be applied on the clusters before matching them with valid indices. Although these two methods can create additional clusters without matching indices, our decoder is able to filter the nonvalid clusters.

Supplementary Table 6 shows that for Illumina test dataset the failure rate of the DNAformer achieves almost the same results as the perfect clustering algorithm and the data can be retrieved by our decoder. Furthermore, in the Nanopore test dataset single flowcell, the failure rate when using the binning method is higher compared to the perfect clustering and the Hash-based clustering[22], however in all the three cases it is possible to retrieve the data. Finally, for the Nanopore dataset with two flowcells the DNAformer failure rate is lower when using the binning algorithm compared to the perfect clustering. The Clover-based clustering method[21] does not allow for successful retrieval of information on our datasets.

| | Test dataset \| Illumina | | | Test dataset \| Nanopore single flowcell | | | Test dataset \| Nanopore two flowcells | | |
|---|---|---|---|---|---|---|---|---|---|
| Clustering method | Runtime (min) | No. of clusters | Failure rate | Runtime (min) | No. of clusters | Failure rate | Runtime (min) | No of clusters | Failure rate |
| Binning | 0.4 | 109,944 | 0.055% | 0.36 | 109,928 | 4.79% | 0.483 | 109,976 | 2.01% |
| Perfect | 14,434 | 109,948 | 0.053% | 12,232 | 109,970 | 3.8% | 22,309 | 109,970 | 3.18% |
| Hash[22] | 94 | 491,309 | 28.4% | 124 | 218,681 | 3.05% | 603 | 508,605 | 1.32% |
| Clover[21] | 4 | 97,717 | 85.49% | 8.54 | 299,190 | 44.34% | 14.44 | 368,027 | 35.48% |

**Supplementary Table 6 | Comparison of different clustering methods.** The results show that our binning method, although very simple and fast, achieves competitive failure rates with other methods. Showcasing how other components in the retrieval pipeline can operate with the additional noise. The results also show that although the Hash method creates more clusters than the original file contained, our decoder is able to filter the incorrect clusters and retrieve the information on the Nanopore datasets. For all clustering methods examined, the reconstruction was done with the DNAformer.

# Reconstruction

## Model architecture

The DNAformer architecture follows a Siamese structure with two branches and shared weights fused together to form a single unified predicted sequence, as shown in Fig. 1b. First an alignment module is applied whose purpose is to learn the required alignment for each read independently. The alignment module uses an Xception[48] inspired architecture with depthwise separable 1D convolutions and multiple kernel heads with size of 1, 3, 5 and 7. The purpose of using multiple kernels in the embedding layer is to allow the model to capture different shifts caused by deletion or insertion errors. The module is constructed with a repeatable block of a linear layer, followed by layer normalization and GELU activation. In addition, this module outputs a sequence with the required output length.

Following this module, we sum over the cluster dimension with the goal of improving the model's robustness to differences in cluster size. This operator has the effect of NCI and aims to increase the SNR of the data.

After the NCI aligner layer, the architecture employs an embedding module whose architecture is similar to the alignment module, however, now the operations are employed on the whole cluster and not on each read independently. The goal is to learn correlations between the different reads and prepare the data for the Transformer layer.

The transformer layer is a multi-head transformer architecture with linear layers for the feedforward part. We do not use position embeddings in this module. After the last transformer block, a linear module is used to reduce the number of features to 4 which represent one-hot encoding for DNA representation.

The fusion vector has the length of the encoded sequence, and its purpose is to learn the optimal combination between the two branches. The fusion vector parameters are initialized with a constant value of 1 for each index, which means equal contribution from each branch. The results provided in Supplementary Fig. 5 show different behavior for each experiment reflecting that the model learns different optimal combinations for different sequencing technologies and noise patterns.

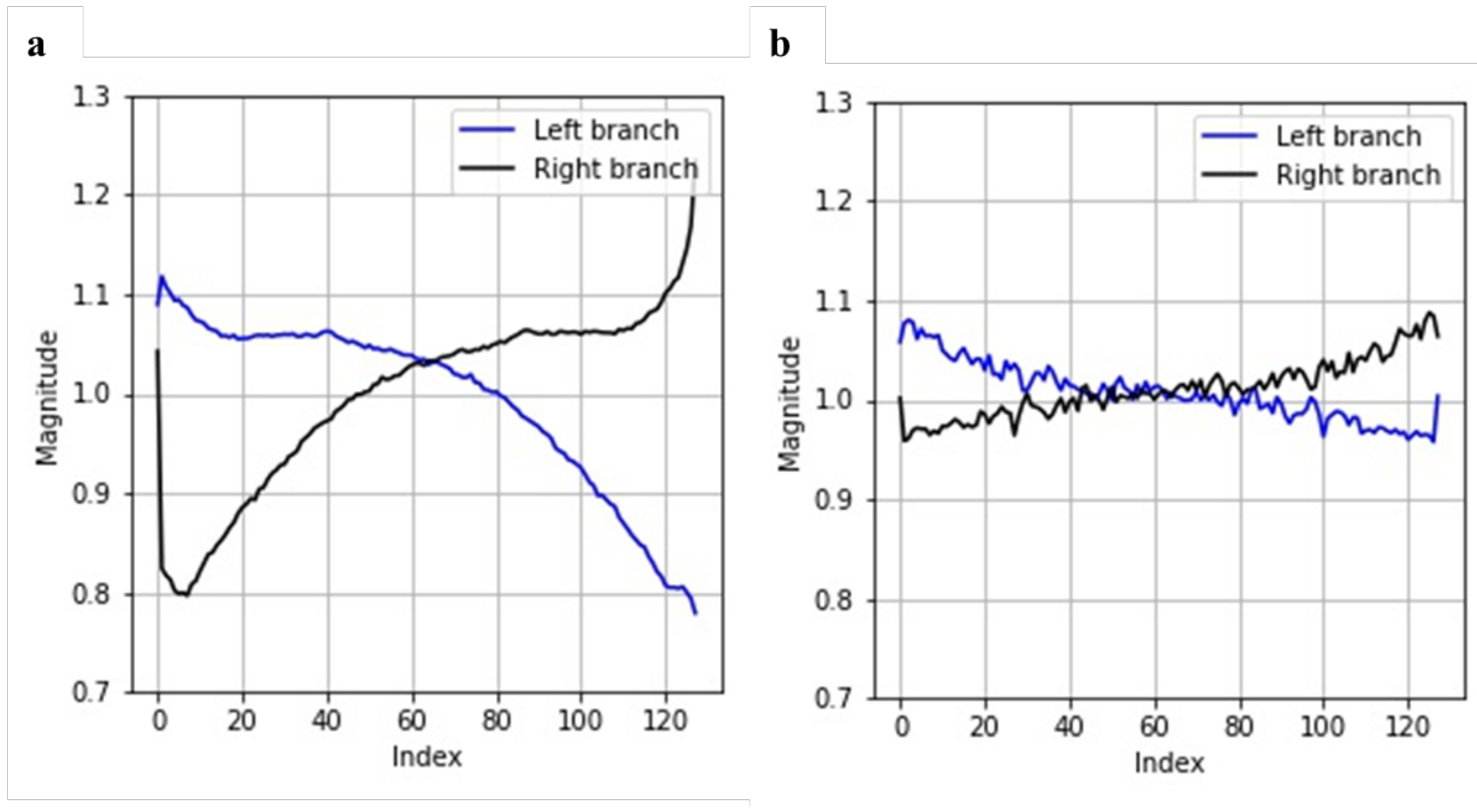

**Supplementary Fig. 5 | Fusion vector values. a** Nanopore experiment. **b** Illumina experiment. The fusion vector learns to combine between the left and right branches in the DNN architecture and yields a single prediction sequence with the length of the required encoded sequence.

Supplementary Table 7 shows the results of the ablation study conducted on the DNAformer architecture using the Nanopore test two flowcells dataset. The results show performance improvement as model capacity increases, coinciding with results of DNNs for language and computer vision tasks. Additionally, we see that DNAformer trained on the pilot dataset statistics is able to achieve reconstruction of the full dataset. The Siamese architecture allows to introduce an additional loss mechanism to enforce consistency between the left and right branches and shows improvement other using cross-entropy only.
Several data augmentations methods were examined. The first mechanism injects random false copies into the training process. This helps the model to learn how to ignore such cases which occur due to clustering errors. The second mechanism varies the stand deviation of the generated noise statistics in the data generator during training. Lastly, the third mechanism controls each type of error (substitution, insertion, or deletion) separately to vary the generation process even further.

| Architecture | Model capacity | Data generator | Loss | Augmentations | | | | | No. of errors |
|---|---|---|---|---|---|---|---|---|---|
| | | | | Substitution deviation | Insertion deviation | Deletion deviation | STD deviation | False copies | |
| Single branch | 20M | Nanopore test | CE | Off | Off | Off | Off | Off | 8456 |
| Single branch | 70M | Nanopore test | CE | Off | Off | Off | Off | Off | 4825 |
| Single branch | 70M | Nanopore test | CE | Off | Off | Off | 0.1 | Off | 4734 |
| Single branch | 70M | Nanopore test | CE | Off | Off | Off | Off | 2 | 4233 |
| Single branch | 70M | Nanopore test | CE | Off | Off | Off | 0.1 | 2 | 3754 |
| Single branch | 70M | Nanopore test | CE | 75% | 75% | 75% | Off | Off | 5854 |
| Single branch | 70M | Nanopore test | CE | 125% | 125% | 125% | 0 | Off | 4437 |
| Single branch | 100M | Nanopore pilot | CE | Off | Off | Off | 0.1 | 2 | 3362 |
| Siamese | 100M | Nanopore pilot | CE + Consistency | Off | Off | Off | 0.1 | 2 | 2896 |
| Siamese | 100M | Nanopore pilot | CE + Consistency | 125% | 125% | 125% | 0.1 | 2 | 2720 |
| Siamese + CPL | 100M | Nanopore pilot | CE + Consistency | 125% | 125% | 125% | 0.1 | 2 | 1842 |

**Supplementary Table 7 | DNAformer ablation study.** In this section we focused on the Nanopore test two flowcells dataset. The examination included single branch and Siamese architecture where we see a clear benefit of using the latter. In addition, this architecture also allowed us to incorporate a consistency loss in parallel to the CE loss. From the results reported it is evident that the increase in model size also improves the results. Several data augmentation strategies were examined, and their collective contribution is reported. Lastly, combining the DNN with the CPL improves the results further, as additional small clusters are solved by the CPL.

## Simulated data generation

In our comparison on publicly available datasets, we model the noise characteristics based on the dataset. Therefore, it is important to verify that there is no case of overfitting when reporting the results on these datasets. To verify this issue, we randomly split each dataset in half. One half of each dataset was used to generate the error statistics for our data generator and the other half was used for evaluation. In other words, the evaluation was performed on un-modelled data.

The results of this examination are presented in Supplementary Table 8, where we used the single branch 70M parameters configuration for fast experimentation. We report a good fit between the evaluation on the full and split datasets, thereby supporting the results reported in Fig. 3.

| | Erlich et al.[14] | | Grass et al.[12] | | Organick et al.[15] | |
|---|---|---|---|---|---|---|
| Dataset | Full | Split | Full | Split | Full | Split |
| Synthesis | Twist Bioscience | | CustomArray | | Twist Bioscience | |
| Sequencing | Illumina miSeq | | Illumina miSeq | | Illumina NextSeq | |
| Dataset size | 72,000 | 36,001 | 4,989 | 2,495 | 596,499 | 298,249 |
| Number of wrong predictions | 16 | 7 | 64 | 31 | 1,373 | 630 |
| Error percentage | 0.02% | 0.019% | 1.3% | 1.25% | 0.23% | 0.21% |

**Supplementary Table 8 | Performance comparison between full and split datasets.** The results show a good fit between the full and split dataset cases. The split dataset case used half of the data to generate error statistics for the data generator and evaluated on the other half of un-modelled data. Wrong prediction is defined as having at least one wrong character out of the entire predicted sequence.

Assessment on the relationship between real and simulated data for training was performed using the dataset provided by Organick et al.[15] due to its relatively large size. The results are provided in Supplementary Table 9 where we examined four data configurations. All configurations used the single branch 70M parameters configuration. To create the evaluation, the dataset was split into 425,006 frames for validation and 141,668 frames were used to model the error statistics and were also used to train the 'Real data' configuration. The 'label + simulated data' configuration used real DNA designed strands as labels and generated simulated reads at each iteration. The 'mixed real and simulated data' configuration utilized a linear, progressive blending between the two data sources, which started with simulated data only in the first epoch and ended with real data only in the last epoch. The results show that utilizing our data generator can not only replace real data, which is expensive and time consuming to acquire, but also improve performance, largely due to the un-limited amount of simulated data that can be generated. Furthermore, the results show a small improvement when combining between a small amount of real data and a large amount (x10) of simulated data.

| | Simulated data | Labels + simulated data | Real data | Mixed real and simulated data |
|---|---|---|---|---|
| Number of wrong predictions | 988 | 1155 | 1460 | 948 |
| Failure rate | 0.23% | 0.27% | 0.34% | 0.22% |

**Supplementary Table 9 | Comparison between real and synthetic data performance.** The results show the proposed data generator achieves better performance than real data-based training. The combination of real and simulated data achieved the best performance. Wrong prediction is defined as having at least one wrong character out of the entire predicted sequence.

## The CPL algorithm

In this subsection, we first give a high-level description of the computational steps of the CPL algorithm:

1. **Input.** The algorithm receives a cluster, denoted by $\boldsymbol{C}$ of $\boldsymbol{t}$ reads, and the encoded sequence's length $u$ and estimates the original encoded sequence of $\boldsymbol{C}$.
2. **Edit distance calculation.** The algorithm considers the first read of $\boldsymbol{C}$, denoted by $\boldsymbol{y}_1$ and uses a dynamic programming to calculate the selected reads' edit distance from the any other read in the cluster. This step involves calculation of the edit distance of $t-1$ pairs of reads.
3. **Edit distance calculation.** The algorithm uses a backtracking technique on the edit distance calculation to retrieve edit-operation vectors that describe how to change $\boldsymbol{y}_1$ to any of the other reads in the cluster.
4. **Edit distance graph generation.** The algorithm utilizes the set of edit-operation vectors to form a graph, in which the vertices set consists of two types of vertices; The first type corresponds to the symbols and their locations as they appear in the selected read, while the second type is correlated with possible insertion operations between the symbols of the selected read. The edges connect between consecutive (by index) vertices of the first type and adjacent vertices of the two types related to an insertion operation. The weights of the edges are defined as the logarithm of the estimated probability of observing the edge and its two consecutive vertices as part of a path that corresponds to the algorithm's output.
5. **Longest path calculation.** Finally, the longest path in the graph is computed, which corresponds to a sequence of length $u$, and returns its corresponding sequence as the algorithm's output estimation.

Detailed description of the CPL algorithm is provided below.

**Input and edit distance calculations.** The algorithm chooses the first reads in the cluster and denotes it by $\boldsymbol{y}_1$. Using a dynamic programming technique[52], we calculate the edit distance between $y_1$ and each of the other reads in the cluster $\boldsymbol{y}_k$ for $2 \leq k \leq t$. To compute the edit distance, the algorithm creates a $|\boldsymbol{y}_1| \times |\boldsymbol{y}_k|$ dynamic programming matrix, in which the $(i,j)$-th cell of the matrix represents the edit distance between the first $i$ symbols in $\boldsymbol{y}_1$ and the first $j$ symbols in $\boldsymbol{y}_k$. Next, the algorithm uses a backtracking method on the computed matrices to retrieve for each pair $\boldsymbol{y}_1$ and $\boldsymbol{y}_k$ a vector of edit operations that describes how to obtain $\boldsymbol{y}_k$ from $\boldsymbol{y}_1$. This vector is denoted by $\boldsymbol{EV}(\boldsymbol{y}_1, \boldsymbol{y}_k)$ and the set of all such vectors is denoted by $\boldsymbol{EV}_{\boldsymbol{y}_1}(\boldsymbol{C})$.

**Edit-operations vectors preparation.** There are four possible operations: copy, insertion, deletion, and substitution. Note that insertion operations can cause various lengths for the vectors in $\boldsymbol{EV}_{\boldsymbol{y}_1}(\boldsymbol{C})$, and thus to avoid this confusion, we partition each edit-operation vector into two vectors of length $|\boldsymbol{y}_1|$ and $|\boldsymbol{y}_1|+1$. The first vector is denoted by $\boldsymbol{EV}_{\boldsymbol{CDS}}(\boldsymbol{y}_1, \boldsymbol{y}_k)$, and it corresponds to copy, substitution, deletions operations. That is, for $1 \leq i \leq |\boldsymbol{y}_1|$, the $i$-th entry of the vector $\boldsymbol{EV}_{\boldsymbol{CDS}}(\boldsymbol{y}_1, \boldsymbol{y}_k)$ describes whether the $i$-th symbol of $\boldsymbol{y}_1$ should be copied, deleted, or substituted with another symbol in order to transform $\boldsymbol{y}_1$to $\boldsymbol{y}_k$. The second vector is denoted by $\boldsymbol{EV}_{\boldsymbol{I}}(\boldsymbol{y}_1, \boldsymbol{y}_k)$ and it describes the required insertion operations that need to be performed on $y_1$ to transform it to $\boldsymbol{y}_k$. That is, for $1 \leq i \leq |\boldsymbol{y}_1|$, the $i$-th entry of $\boldsymbol{EV}_{\boldsymbol{I}}(\boldsymbol{y}_1, \boldsymbol{y}_k)$ describes the symbol(s) that should be inserted to $\boldsymbol{y}_1$ before its $i$-th symbol to transform it to $\boldsymbol{y}_k$. For simplicity, the possible symbols also include the empty word $\varepsilon$, which corresponds to the case where no insertion is required. It should be mentioned that for given $\boldsymbol{y}_1$ and $\boldsymbol{y}_k$, the vector $\boldsymbol{EV}(\boldsymbol{y}_1, \boldsymbol{y}_k)$ is not necessarily unique, and in this case the algorithm chooses one of them randomly.

**Example of edit-distance calculation and edit-operations vectors**. Supplementary Fig. 6 provides an example which describes the dynamic programing matrix for two possible reads $\boldsymbol{y}_1 = ATTACA$ and $\boldsymbol{y}_k = ATGCTG$.

| - | - | A | T | T | A | C | A |
|---|---|---|---|---|---|---|---|
| - | 0 | 1 | 2 | 3 | 4 | 5 | 6 |
| A | 1 | 0 | 1 | 2 | 3 | 4 | 5 |
| T | 2 | 1 | 0 | 1 | 2 | 3 | 4 |
| G | 3 | 2 | 1 | 1 | 2 | 3 | 4 |
| C | 4 | 3 | 2 | 2 | 2 | 2 | 3 |
| T | 5 | 4 | 3 | 2 | 3 | 3 | 3 |
| G | 6 | 5 | 4 | 4 | 3 | 3 | 3 |

**Supplementary Fig. 6 | Example of edit-distance calculation and edit-operations vectors**. In this example of a dynamic programing matrix, $\boldsymbol{y}_1 = ATTACA$ and $\boldsymbol{y}_k = ATGCTG$.

By backtracking the matrix described in Supplementary Fig. 6, we can derive the following two error vectors as defined above:

$$EV_{CDS}(\boldsymbol{y_1}, \boldsymbol{y_k}) = \big((0, copy, A), (1, copy, T), (2, del, T), (3, sub, A \to G), (4, copy, C), (5, copy, A)\big)$$

$$EV_I(\boldsymbol{y_1}, \boldsymbol{y_k}) = \big((0, ins, \varepsilon), (1, ins, \varepsilon), (2, ins, \varepsilon), (3, ins, \varepsilon), (4, ins, \varepsilon), (5, ins, \varepsilon), (6, ins, G)\big)$$

**Estimations of the error probabilities.** Next, the algorithm uses the set of error vectors $\boldsymbol{EV}_{\boldsymbol{y_1}}(\boldsymbol{C})$ to estimate the probability to have a specific symbol in specific index in the algorithm's estimation. For each index $1 \le i \le n$ and for any two symbols $w, v \in \{A, C, G, T\} \cup \{\emptyset\} \cup \{\$\}$ we define the conditional probability to obtain symbol $v$ in index $i + 1$, given that the $i$-th symbol is $w$. The symbol in the 0-th index is strictly defined as $\$ \notin \{A, C, G, T\}$ to mark that this is the beginning of the word, and the symbol $\emptyset$ corresponds to the empty word. The estimated conditional probability is given in the expression: $P_i^{y_j, \boldsymbol{C}}(v|w) \triangleq \frac{\left|\left\{\boldsymbol{v} \in EV_{\boldsymbol{y_j}}(\boldsymbol{C}) : \boldsymbol{v}[i+1] = v \text{ and } \boldsymbol{v}[i] = w\right\}\right|}{\left|\left\{\boldsymbol{v} \in EV_{\boldsymbol{y_j}}(\boldsymbol{C}) : \boldsymbol{v}[i] = v\right\}\right|}$, where $\boldsymbol{v}[i]$ refers to the $i$-th symbol of the vector $\boldsymbol{v}$. Following that, we calculate the set: $Post_{i,w}^{y_j, \boldsymbol{C}} \triangleq \{\, v : v \in EV_{\boldsymbol{y_j}}(\boldsymbol{C})[i+1], P_i^{y_j, \boldsymbol{C}}(v|w) \neq 0\}$, which is the set of the symbols that can be achieved on the $(i + 1)$-th index of an error vector in $EV_{\boldsymbol{y_j}}(\boldsymbol{C})$, given that the $i$-th symbol was $w$. Note that, the set $Post_{i,w}^{y_j, \boldsymbol{C}}$ can be empty.

**Graph generation.** Based on these estimated probabilities, we can now define the edit graph $G_{\boldsymbol{y_j}\boldsymbol{C}} = \left(\left(V_{\boldsymbol{y_j}}, E_{\boldsymbol{y_j}}\right)\right)$ of a read $\boldsymbol{y_j}$ and a cluster $\boldsymbol{C}$. The edit graph of the sequence is defined as follows.

1. $V(\boldsymbol{y_j}) = \left\{w : w \in EV_{\boldsymbol{y_j}}(\boldsymbol{C})[i], 1 \le i \le 2\left|\boldsymbol{y}_j\right| + 1, w \neq \emptyset\right\} \cup \{\boldsymbol{S}\}$. The vertices are defined as the (non-empty word) symbols from the sets $EV_{\boldsymbol{y_j}}(\boldsymbol{C})$. Note that, the same symbol can appear as more than one vertex, depends on its indices in the vectors of $EV_{\boldsymbol{y_j}}(\boldsymbol{C})$. Furthermore, in odd indices of vectors in $EV_{\boldsymbol{y_j}}(\boldsymbol{C})$, there can be more than one symbol. In this case we have a vertex for each symbol that appears in this position of the vector, where the symbols are connected to each other based on their order, where the weight of their edges are 0.

2. $E(\boldsymbol{y_j}) = \{(w, v)_i : w \in EV_{\boldsymbol{y}_j}(\boldsymbol{C})[i], v \in Post_{i,w}^{\boldsymbol{yj},\boldsymbol{C}}, v \neq 0\}$. Each edge connects between any two vertices that represent symbols that appear consecutively in an error vector in the set $EV_{\boldsymbol{y}_j}(\boldsymbol{C})$. In case that $u$ or $v$ consist of more than one symbol, we connect the last symbol of $u$ to the first symbol of $v$.
3. $W: E(\boldsymbol{y_j}) \rightarrow \mathbb{R}$ is a weight function, defined on the edges. The weight of an edge $(w, v)_i \in E(\boldsymbol{y_j})$ is defined as the log of its conditional probability, that is $W((w, v)_i) = \log\left(P_i^{\boldsymbol{y}_j,\boldsymbol{C}}(v|w)\right)$.
4. For $w \in EV_{\boldsymbol{y}_j}(\boldsymbol{C})[i]$ where $\emptyset \in Post_{i,w}^{\boldsymbol{y}_j,\boldsymbol{C}}$, we connect the last symbol of $w$ to the first symbol of any $v \in Post_{i+1,\emptyset}^{\boldsymbol{y}_j,\boldsymbol{C}}, v \neq \emptyset$. The weight of this edge is $log(P_i(\emptyset|w))$+ $log(P_i(v|\emptyset))$. The same holds for any index $j \geq i$, in this case, the last symbol of $w$ is connected to the first nonempty symbol of any $v \in Post_{j,\emptyset}^{\boldsymbol{y}_j,\boldsymbol{C}}, v \neq \emptyset$. In this case, the weight of the edge is defined as $log(P_i(\emptyset|w))$+ $log(P_i(\emptyset|\emptyset)) + \cdots + log\left(P_{j-1}(\emptyset|\emptyset)\right) + log\left(P_j(v|\emptyset)\right)$.
5. The vertex represents the beginning of the sequence, and is connected to any vertex $v$ that corresponds to index $i = 1$ in the error vectors in $EV_{\boldsymbol{y}_j}(\boldsymbol{C})$, with edge of weight exactly $log(P_1(v|\$))$.

**Output of the algorithm.** Finally, the CPL algorithm calculates the heaviest path in the edit graph $G_{\boldsymbol{y}_j\boldsymbol{C}} = \left(\left(V_{\boldsymbol{y}_j}, E_{\boldsymbol{y}_j}\right)\right)$ of length $u + 1$. If there does not exist a path of length $u + 1$, the algorithm picks the one that is closest in length to $u + 1$ edges. If there is more than one, it picks one of them randomly. We denote the number of edges in the selected path as $u^{\sim}$.The $u^{\sim}$ vertices in this path correspond to $u^{\sim} - 1$ symbols and one vertex of the symbol \$. Therefore, the path induces a sequence $\widehat{\boldsymbol{x}}$ that estimates the original sequence of the cluster. The output of the algorithm is $\widehat{\boldsymbol{x}}$. The complexity of the algorithm is $O(tu^2)$ when $t$ is the number of reads and $u$ is the encoded sequence's length.

**Example of the edit graph.**
Supplementary Fig. 7 depicts an example of the edit graph that is used by the CPL algorithm. In this example, we describe a possible input to the CPL algorithm and its corresponding graph. We assume a read which is denoted by $\boldsymbol{y}_j \in \boldsymbol{C}$ is given to the CPL algorithm. The example describes only five positions in $\boldsymbol{y}_j$, whose indices are given by $i, i + 1, i + 2, i + 3, i + 4$. For each of these positions, the specific symbol of $\boldsymbol{y}_j$ is given, together with the observed symbols in the set $Post_{i,w}^{\boldsymbol{y}_j,\boldsymbol{C}}$. The set $Post_{i,w}^{\boldsymbol{y}_j,\boldsymbol{C}}$, in each of the described positions is calculated by the CPL algorithm. Finally, the graph which corresponds to the described input and to the sets $Post_{i,w}^{\boldsymbol{y}_j,\boldsymbol{C}}$ is depcited.

**a**

| Position $i$ | | Position $i+1$ | | Position $i+2$ | | Position $i+3$ | | Position $i+4$ | |
|---|---|---|---|---|---|---|---|---|---|
| *String* | *Post* | *String* | *Post* | *String* | *Post* | *String* | *Post* | *String* | *Post* |
| $A$ | $\{GTC, \emptyset\}$ | $GTC$ | $T$ | $T$ | $C$ | $C$ | $A$ | $A$ | $\emptyset$ |
| | | $\emptyset$ | $\{T, \emptyset\}$ | $\emptyset$ | $\{C, \emptyset\}$ | $\emptyset$ | $A$ | | |

**b**

Position $i$ — Position $i+1$ — Position $i+2$ — Position $i+3$ — Position $i+4$

A, G, T, C, T, C, A

$log(0.5)$, 0, 0, 0, 0, 0, $2log(0.5)$, $3log(0.5)$, $3log(0.5)$

**Supplementary Fig. 7 | Example of the graph of the CPL algorithm**. **a,** example of an input read $\boldsymbol{y}_j \in \boldsymbol{C}$ of the CPL algorithm. Each column corresponds to specific position of the input $\boldsymbol{y}_j \in \boldsymbol{C}$, where the positions are indexed by $i, i+1, i+2, i+3, i+4$. For each of these positions, there first column presents the observed string, denoted by $String\ i$, and the second column presents the values of the set $Post_{i,w}^{yj,C}$ for $w \in String\ i$, which is the set of possible sequences that are adjacent (from right) to the symbols that appear in the $i$-th position. The set $Post_{i,w}^{yj,C}$ is calculated as defined in the CPL algorithm. **b,** shows the edit graph based on the table in **a**.

# Coding scheme

Previous works in DNA-based storage have primarily relied on either inner codes, outer codes, or a combination of both. Inner codes involve encoding each sequence separately using an ECC, which can overcome up to $d$ errors at the nucleotide level. Outer codes, on the other hand, involve encoding the sequences collectively to overcome missing or corrupted encoded sequences by adding redundant ones. However, adding redundancy for each encoded sequence can result in suboptimal use of redundancy, especially when $d$ is chosen to address the worst-case scenario of reads. Conversely, if $d$ is too small, many encoded sequences may be decoded incorrectly, requiring a high level of redundancy in the outer code to correct them. To overcome this tradeoff and utilize the inherent redundancy more effectively, we propose using a TP based coding scheme that can correct up to $d$ errors in up to a fraction $p$ of the encoded sequences. This allows a significant reduction in the number of redundancy symbols used for error correction while maintaining error free retrieval of information.

Our suggested coding scheme encodes a block of $b$ bits into $M = 55{,}000$ encoded sequences of length $L = 140$. The parameter $b$ depends on the specific setup in which our constrained code is applied. Here we propose two different setups, the first enforces the length of the longest homopolymer to be at most 4 and the GC content of each encoded sequence to be between 30% to 70%. The second setup does not enforce any constraint on the encoded sequences. For the first setup, the total number of information bits is $b = 12{,}317{,}012$ bits (1.53963MB) and for the second setup $b = 13{,}249{,}024$ bits (1.656128MB).

Each strand is composed of 12 bases for index, including a single base that serves as a file identifier and $u$ =128 bases for data and redundancy as presented in Supplementary Fig. 8f. A high-level description of our coding scheme is given in Supplementary Fig. 8a-e. Our coding scheme is based on the following components:

1. **Index encoding**. A mapping function $E_{Ind}: [M] \to \{A, C, G, T\}^{11}$ which is used to create the indices for the encoded sequences, i.e., for any $i \in \{0, \dots, M-1\}$, the index of the $i$-th encoded sequence is $E_{Ind}(i)$. Here we designed the mapping $E_{Ind}$ such that the indices of the different encoded sequences are far enough from each other and satisfy the constraint that the length of every homopolymer is at most 4. The index itself is of length $L - u = 12$. The additional base, which is the file identifier, is added later to satisfy some structural constraints. More details can be found under the Index encoding subsection.
2. **Diagonal-column encoding**. A mapping $E_{DRS}: \{0,1\}^{(M-r_1)\times 16\cdot 13} \to \{0,1\}^{M\times 16\cdot 13}$. The mapping $E_{DRS}$ encodes the columns of a binary $(M - r_1) \times 16$ matrix diagonally to protect from any remaining errors that were not corrected using the information retrieval process or the other coding components. This code serves as a fail-safe mechanism, and by designing other components carefully, its redundancy, $r_1$, can be relatively low. For more details see the Diagonal-column encoding subsection.
3. **Constrained code**. A mapping function $E_{cons}: \{0,1\}^{n_1} \to \{A, C, G, T\}^{n_2}$ such that $\frac{n_1}{2} \leq n_2 < n_1$. The mapping $E_{Cons}$ encodes a binary sequence of length $n_1$ into a length $n_2$ sequence over $\{A, C, G, T\}$. In this work, we suggest a possible mapping that satisfies two constraints: (1) GC-content between 45% and 55% with high probability and between 30% and 70% in the worst case, and (2) homopolymers of length 4 or less. For more details see the Constrained code subsection.
4. **Tensor-product code**. Our method utilizes a TP code, which involves a Bose–Chaudhuri–Hocquenghem (BCH) code[13] with redundancy $r_2$ and a RS code with redundancy $r_3$. Our TP code can be described as a mapping $E_{TP}: \mathcal{M}(M, u, r_2, r_3): \to \{A, C, G, T\}^{M\times u}$ such that for $r_2 \leq M$, and $r_3 \leq u$, $\mathcal{M}(M, u, r_2, r_3)$ is the set of all $M \times u$ matrices over $\{A, C, G, T\}$ in which the bottom-right $r_3 \times r_2$ sub-matrix is empty. For more details about the applicable parameters and encoding, and decoding algorithms for $E_{TP}$ see Tensor-product code and Decoding subsections.

## Encoding description

Our encoding scheme encodes a binary data $\boldsymbol{x}$ into a codeword matrix $X \in \{A, C, G, T\}^{M\times L}$. The redundancy parameters that were used in our dataset are $r_1$=3,026, $r_2$=16, and $r_3$=4,786. The number of bits in $\boldsymbol{x}$ is denoted by $b$, where $b = 238 \cdot (M - r_3) + 208 \cdot (r_3 - r_1)$ in our constrained setup, and $b = 256 \cdot (M - r_3) + 224 \cdot (r_3 - r_1)$ in our non-constrained setup. In the description to follow we use the notation $b = b_\ell \cdot (M - r_3) + b_s \cdot (r_3 - r_1)$ where $b_\ell$ and $b_s$ should be set according to the desired constrained setup. The encoding of $\boldsymbol{x} \in \{0,1\}^b$ into $X \in \{A, C, G, T\}^{M\times L}$ is done by the following steps.

1. **Index encoding.**
   1.1. Define an $M \times (L - u)$ matrix $\mathcal{I}$, where for any $i \in [M]$, fill the $i$-th row of $\mathcal{I}$ with the index $E_{Ind}(i)$.
   1.2. Create an empty $M \times b_\ell$ matrix $\mathcal{X}_b$ and concatenate it to the right of the matrix $\mathcal{I}$. The output of this step is depicted in Supplementary Fig. 8a.
2. **Input parsing.**
   2.1. Fill the matrix $\mathcal{X}_b$ with the bits of $\boldsymbol{x}$, such that the first $M - r_3$ rows contain exactly $b_\ell$ bits each, and the next $r_3 - r_1$ rows contain $b_s$ bits each (the rest of the matrix remains empty).
   2.2. Denote by $A', A''$ and $B'$ the following submatrices of $\mathcal{X}_b$.
      - $A'$ is the top-left $(M - r_3) \times b_s$ submatrix of $\mathcal{X}_b$.

- $A''$ is the top-right $(M - r_3) \times (b_\ell - b_s)$ submatrix of $\mathcal{X}_b$.
- $B'$ is the $(r_3 - r_1) \times b_s$ submatrix of $\mathcal{X}_b$ located just below $A'$.

This partition of $\mathcal{X}_b$ is depicted in Supplementary Fig. 8b.

3. **Diagonal-columns encoding.**
   Encode the $(M - r_1) \times b_s$ submatrix that is composed of $A'$ and $B'$ using $E_{DRS}$. This step adds $r_1$ redundancy bits to each column and the submatrix that corresponds to these redundancy bits is denoted by $C'$ as depict in Supplementary Fig. 8c.
4. **Constrained code.**
   Encode each of the rows of $\mathcal{X}_b$ using $E_{cons}$. To this end, we use two instances of $E_{cons}$ as follows.
   4.1. Encode the first $M - r_3$ rows (longer rows) we use $E_{cons}$ with parameters $n_1 = b_\ell$ and $n_2 = 128$.
   4.2. Encode the last $r_3$ rows (shorter rows), we use $E_{cons}$ with parameters $n_1 = b_s$ and $n_2 = 128 - r_2$.
   The output of this step is presented in Supplementary Fig. 8d1, where the 4-ary representation of $A', A''$ is denoted by $A$, the 4-ary representation of $B'$ is denoted by $B$, and the 4-ary representation of $C'$ is denoted by $C$.
   4.3. Remove the $b_\ell - 128$ columns of the matrix $\mathcal{X}_b$ to obtain the $M \times u$ matrix $\mathcal{X}$. Note that $\mathcal{X}$ is a matrix over $\{A, C, G, T\}$ of size $M \times u$ in which the right-bottom $r_3 \times r_2$ submatrix is empty; see Figure Supplementary Fig. 8d2.
5. **TP encoding.** Encode the matrix $\mathcal{X}$ using $E_{TP}$ to complete the $r_3 \times r_2$ missing symbols as depict in Supplementary Fig. 8e.

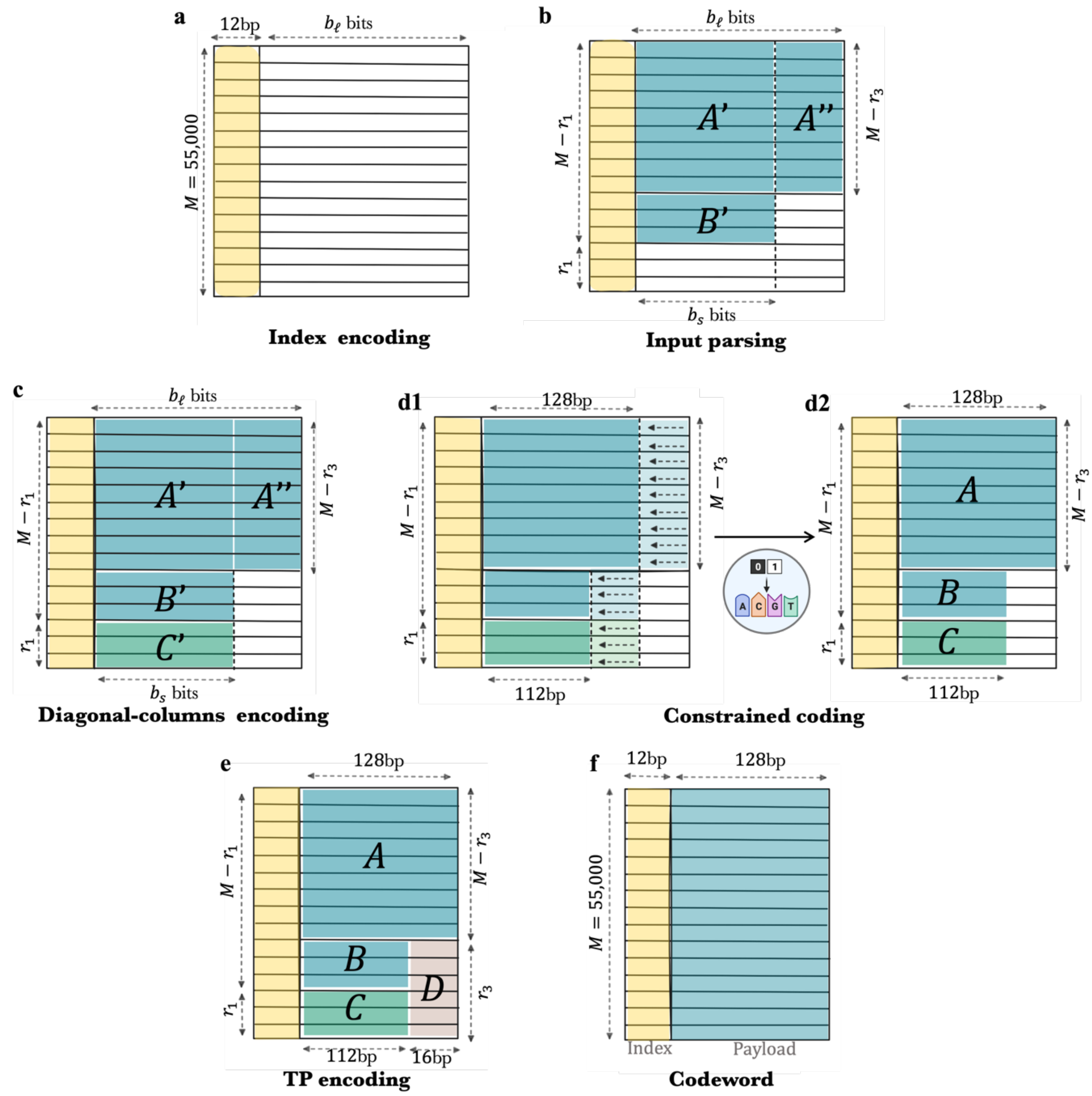

**Supplementary Fig. 8 | Encoding scheme.** The main steps in our encoding scheme are shown. **a,** index encoding. Assigning each row, a unique identifier. **b,** input parsing. Organizing the binary data into a specific structure within a matrix form. **c,** diagonal-columns encoding. Using an RS code in a diagonal manner to encode $A'$ and $B'$. **d1-2,** constrained coding. Convert the binary data into base 4 (corresponding to {A, C, G, T} molecules) while satisfying design constraints. **e,** TP coding. The main component that is responsible for error correcting. Adds additional redundancy symbols to protect A, B, and C matrices against substitutions errors. **f,** codeword. The output of our coding scheme, each row serves as an encoded sequence.

## Index encoding

The set of $M = 55{,}000$ indices of length 11 were selected randomly from the set of all sequences of length 11 over $\Sigma = \{A, C, G, T\}$, such that the maximal homopolymer's length is 4. To further demonstrate the robustness of our method, we used the same set of $M = 55{,}000$ indices for both files. To distinguish between the two files, we extended the index with an additional base, that serves as a file identifier (which is the 12-th symbol of the index). The file identifier for File 1 is either $A$ or $C$, while the file identifier for File 2 is either $G$ or $T$. By using the set of indices for the two files, we deliberately created a small number of pairs of indices with Hamming and edit distance equal to one. The results presented in this work show that our information retrieval pipeline allows for a small number of close indices which makes the design of the indices a much simpler task.
The exact file identifier is selected such that the homopolymer constraint is maintained (for more details see the Constrained code subsection). The set of $M = 55{,}000$ indices of length 11 and the set of 110,000 indices of length 12 that were used for File 1 and File 2 can be found in the code repository published with this work.

Supplementary Fig. 9 presents a histogram of the edit and Hamming distances between pairs of indices in our set of indices. The results for the $M = 55{,}000$ indices of length 11 are given in Supplementary Fig. 9a and the results for the 110,000 indices that were used in our dataset are given in Supplementary Fig. 9b. Even though we selected our indices randomly, most of the pairs are at Hamming and edit distance at least 5 from each other which is sufficient for our binning algorithm. This behavior was obtained by selecting the length of the indices to be sufficiently large with respect to the number of indices our scheme requires.

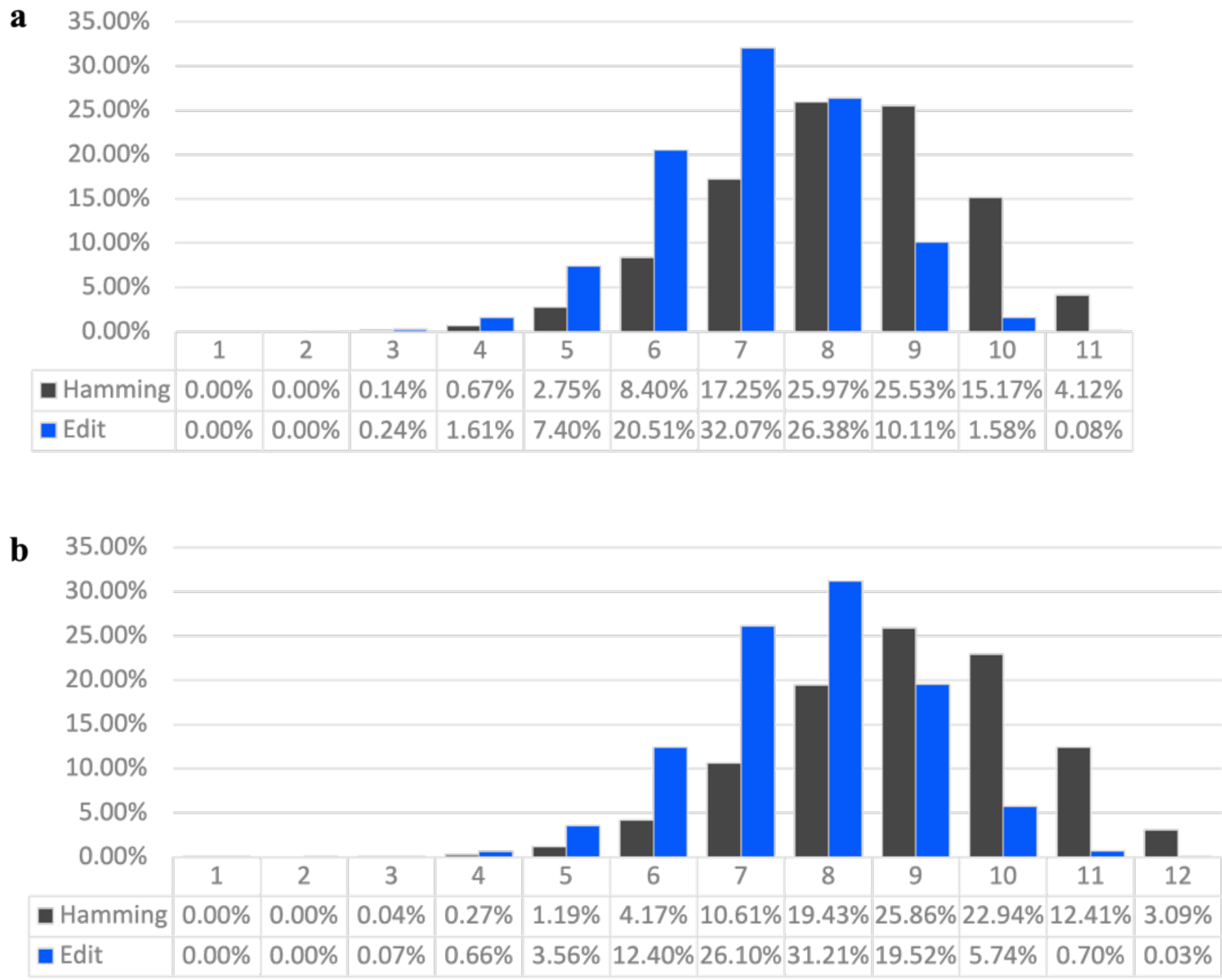

| a | 1 | 2 | 3 | 4 | 5 | 6 | 7 | 8 | 9 | 10 | 11 |
|---|---|---|---|---|---|---|---|---|---|---|---|
| Hamming | 0.00% | 0.00% | 0.14% | 0.67% | 2.75% | 8.40% | 17.25% | 25.97% | 25.53% | 15.17% | 4.12% |
| Edit | 0.00% | 0.00% | 0.24% | 1.61% | 7.40% | 20.51% | 32.07% | 26.38% | 10.11% | 1.58% | 0.08% |

| b | 1 | 2 | 3 | 4 | 5 | 6 | 7 | 8 | 9 | 10 | 11 | 12 |
|---|---|---|---|---|---|---|---|---|---|---|---|---|
| Hamming | 0.00% | 0.00% | 0.04% | 0.27% | 1.19% | 4.17% | 10.61% | 19.43% | 25.86% | 22.94% | 12.41% | 3.09% |
| Edit | 0.00% | 0.00% | 0.07% | 0.66% | 3.56% | 12.40% | 26.10% | 31.21% | 19.52% | 5.74% | 0.70% | 0.03% |

**Supplementary Fig. 9 | Analysis of the edit and Hamming distances of the indices set. a,** shows the results for indices of length 11. **b,** shows the results for indices of length 12, including the file identifier. The histogram shows that most of the indices are at a distance of 5 or more from one another.

## Diagonal columns encoding

Similarly, the methods that have been presented in previous works[12,14,19] in step 3 of our encoding process, we encode the column of our matrix with an error-correcting code, more specifically an $(M, M - r_1, r_1 + 1)$ RS code.
As some sequencing and synthesis technologies are more error-prone at the beginning and end of each read[39], designing the RS code to work directly on the columns of our matrix would have required more redundancy symbols in the encoding of the columns closer to the beginning and end of the encoded sequences (left and right sides of the matrix). Hence, we encode columns diagonally to allow more uniform distribution of the errors. This allows us to use the same amount of redundancy symbols for each diagonal column. Our diagonal columns encoding is presented in Supplementary Fig 13 and is performed as follows. Note that the figure represents the case in which we encode using constraint coding. In the non-constrained setup, there should be an additional 16-bit column that is used to encode information.

1. Given the $(M - r_1) \times b_s$ binary matrix composed of $A'$ and $B'$, partition the matrix into $b_s/16$ blocks of columns (each row composed of 16 bits) as depict in Supplementary Fig 10a.
2. Define $b_s/16$ diagonal columns by shifting the block that is taking from each row by 16 bit from the location in the previous row as depict in Supplementary Fig 10b.
3. Encode each diagonal column using the encoder of the RS code as depicted in Supplementary Fig 10c.
4. Place the redundancy symbols according to the same diagonal shift as shown in Supplementary Fig 10d.

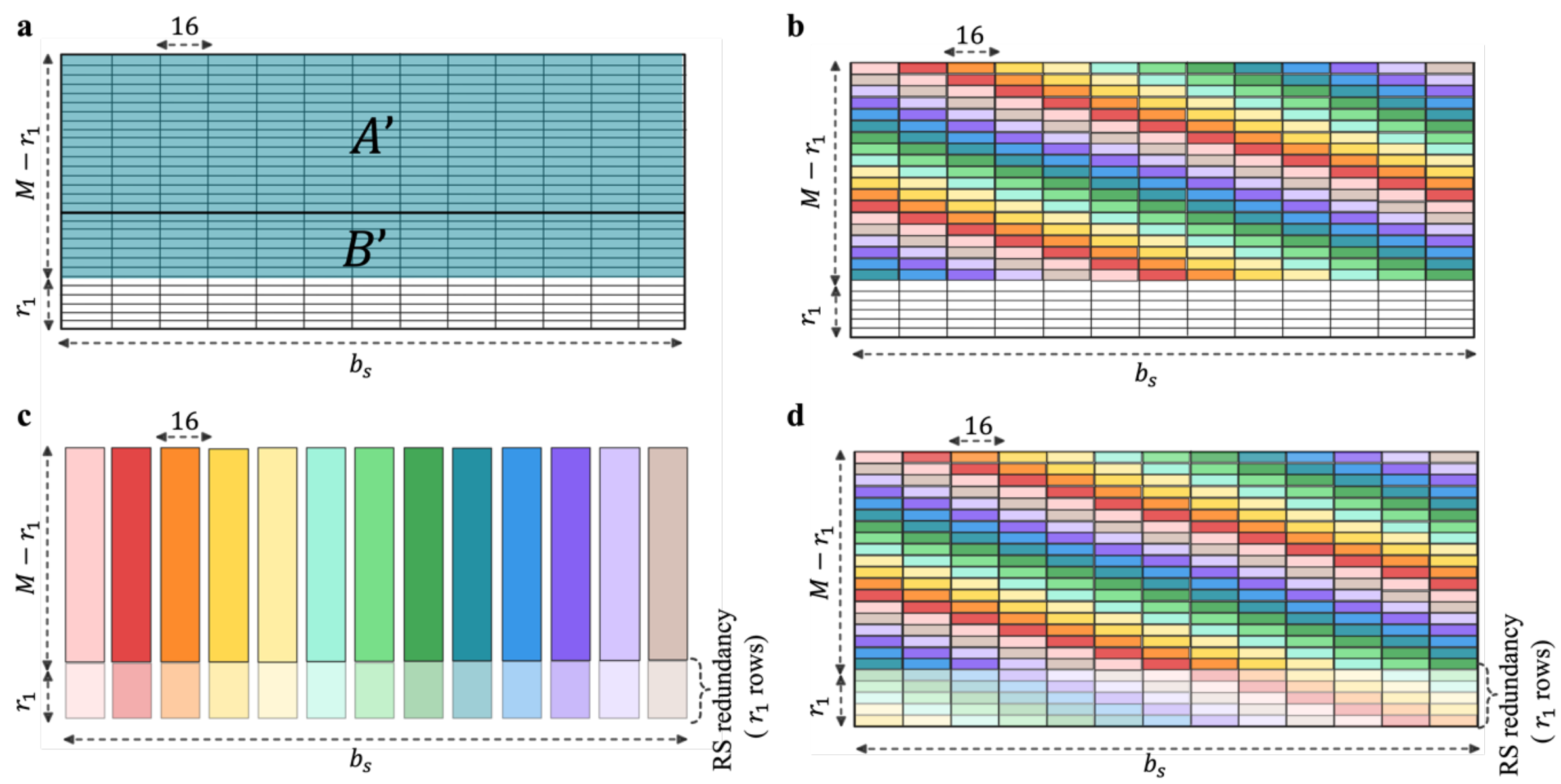

**Supplementary Fig. 10 | Schematic description of the diagonal-column encoding. a,** partition the matrix into 16 bits width columns. **b,** define the diagonal columns by cyclic shift of 16 bits between each row. **c,** encode each diagonal column using RS code. **d,** insert the redundancy from the RS code back into the appropriate diagonal column.

## Constrained code

In DNA-based storage systems there are several structural constraints that should be enforced on the encoded sequences. Such constraints include limiting the length of the homopolymers, avoiding the presence of specific motifs (short sequences) as substrings of the encoded sequences, and controlling the amount of $G$ and $C$ bases that occur in each encoded sequence. Such constraints are used to mitigate the error rates in the biological process (synthesis, PCR, and sequencing) involved in DNA-based storage systems, and the specific constraints and the parameters of these constraints should be selected according to the specific technology that is being used.

Incorporating constrained codes together with ECC requires careful attention as these two modules cannot be concatenated with one another in a straightforward manner as there is no general approach to how to pair the two. To overcome this challenge, we designed our constrained code to interleave with the ECC in a simple manner that allows the selection of specific design constraints. For our experiment, we focused on the constraints of limiting the length of the homopolymers to be at most 4. Additionally, we restrict the amount of $G$ and $C$ bases in each encoded sequence to be between 30% and 70% in the worst case and between 45% to 55% with high probability. We note that the suggested constrained coding is based on block encoding and thus it can be adapted to support additional constraints by designing the exact mapping between a block of bits to a block of DNA bases. The only thing that should be maintained to integrate this modified code into our scheme is that the decoding of each block can be done independently of the other blocks.

Our constrained code is based on a predefined mapping function that takes $n_1$ bits and translates them to a sequence of length $n_2$ over the DNA alphabet $\{A, C, G, T\}$. The encoding is done by partitioning the $n_1$ bits into non-overlapping blocks and encoding the blocks of bits into blocks over the DNA alphabet sequentially. If we don't want to enforce any constraints, we use the simple 2 bits to 1 symbol mapping: $00 \rightarrow A, 01 \rightarrow C, 10 \rightarrow G, 11 \rightarrow T$. Otherwise, $n_1$ can be either $b_s = 208$ or $b_\ell = 238$. In the first case, $n_1 = 208$ and the input is partitioned into 16 blocks of 13 bits each. In the second case, $n_1 = 238$ and the input is partitioned into 18 blocks, the first 16 are of length 13 each, and the last 2 blocks are of length 15 each. The blocks of 13 bits are encoded into DNA blocks of length 7, while the blocks of length 15 are encoded into DNA blocks of length 8. The set of allowed DNA blocks (with the corresponding length) consists of all the sequences over $\{A, C, G, T\}$ that do not contain 5 identical consecutive symbols in the middle and 3 identical consecutive symbols in the edges (which guarantee that concatenation of such blocks will never result with a sequence of 5 identical consecutive symbols).

In our coding scheme, we only consider blocks of length 7 or 8. The set of allowed DNA blocks of length 7 is partitioned into 8 groups based on the number of occurrences of $G$ or $C$ symbols within them (from 0 to 7 GC-content). For any binary input of length 13, the mapping either translates the input to a sequence that belongs to the group with 3 or 4 GC content termed, "almost balanced" GC words, or to a couple of sequences, with two different GC contents.
The set of allowed DNA blocks of length 8 is partitioned into 9 groups based on the number of occurrences of G or C symbols within them (from 0 to 8 GC-content). For any binary input of length 15, the mapping either translates the input to a sequence that belongs to the group with 4 GC content, termed "balanced" GC words, or to a couple of sequences with two different GC contents. The complete mappings that were used in our encoder can be found in code repository published with this work.

Next, we describe how our encoding process enforces the constraints.

1. **Limited homopolymer constraint**. As described above, each of the DNA blocks that we use in our mapping has at most two identical bases at both edges. Furthermore, these blocks have at most four identical bases in the middle. Therefore, concatenating any two blocks cannot create a homopolymer of length 5 or above. Additionally, as described in the Index encoding subsection, the first 11 bases of our selected indices do not contain homopolymers of length 5 or above, and their right-most homopolymer is of length at most 4. Thus, it can be verified that we can always select the file identifier (the 12-th symbol of the index) out of the two possible options in a way that will preserve the homopolymer constraint on the entire encoded sequence.
2. **GC-content constraint**. The encoding process of bits into DNA blocks is done sequentially, from left to right. As described in the mapping included in the code repository published with this work, most of the bit blocks are mapped into DNA blocks with either balanced or almost balanced GC content, whereas the rest of the bit blocks are mapped to two different options of DNA blocks with a non-balanced GC content.

Thus, in every step of the encoding process, if there is more than a single DNA block that can be used, the encoder selects the one that keeps the GC content closer to 50%. It can be verified that this ensures GC content of more than 30% and less than 70% (including the index) in the worst case and with high probability the GC content is more than 45% and less than 55%.

Note that the design of our coding scheme allows us to apply the constrained coding on almost all of the encoded sequences in the worst case. However, a small fraction of the encoded sequences (in our case $r_3 = 4786$ encoded sequences) satisfies the constraints only with high probability. More precisely, we cannot apply our constrained coding on the redundancy part of the tensor product code (sub-matrix D in Supplementary Fig. 8e). Meaning that, with small probability, the suffix of the corresponding encoded sequences might contain longer homopolymers.

## Tensor-product code

In inner-outer code approaches, there is an inner code that protects each encoded sequence that encodes the data from errors within its symbols, as well as an outer code that protects an erasure of a strand. By design, the DNAformer reconstruction eliminates most of the deletions and insertions, and the output has the same length as the encoded sequence. Hence, after this step, we need to only take care of substitution errors, which are easier to solve in terms of coding theory and requires less redundancy.

For each cluster, the DNN produces a soft output that is transferred to the confidence filter, which decides whether to produce an output, activate the CPL, or ignore the cluster. The actual outputs of the DNN can be partitioned into four distinct set:

1. Correct predictions - most of the clusters (roughly 85%-100%, see Fig. 3a).
2. Missing clusters - can be corrected by the outer code easily (requires small redundancy).
3. Wrong predictions with a small number of substitutions (3 or less).
4. Wrong predictions with many substitutions (more than 3).

As the size of the 4-th group is very small (in our experiments, between 0% and 3% of the clusters), it is wasteful to encode each of the sequences with an inner code that can correct many errors. Moreover, it is wasteful to encode all sequences with inner code, even for a small number of errors, as most of the predictions are correct. We use these observations to design a code for the DNN outputs and consider the confidence parameter. The key point is that by using the confidence filter, we can classify the outputs from the 4-th group (with very high accuracy) and ignore them (i.e., classifying them as missing clusters) or activate the CPL algorithm on the corresponding clusters. It should be noted that the tensor-product method can be used without the confidence of the DNN, but in that case, additional redundancy will be required.

Our encoding process utilizes a matrix H, which is a $16 \times 128$ parity-check matrix of a BCH code over the 4-ary alphabet, that can correct 3 substitution errors. Additionally, the matrix **H** can be found in the code repository published with this work. Given the matrix **H**, and the matrices $A$, $B$, and $C$, obtained from the previous steps of our encoding (see Supplementary Fig 11), the TP encoding works as follows.

1. Define an $M \times r_2$ empty matrix $S$.

2. For each of the first $M - r_3$ rows of the matrix $\mathcal{X}$ (the rows that correspond to the sub-matrix $A$), update the i-th row of $S$, $S_i$, with the vector $H \cdot r_i^T$ , where $r_i$ is the i-th row of $A$. This step is presented in Supplementary Fig. 11a.
3. Encode the first $M - r_3$ rows of $S$ using a RS code with $r_3$ redundancy symbols as depicted in Supplementary Fig. 11b.
4. To obtain the matrix $X$ from $\mathcal{X}$, fill the $r_3 \times r_2$ bottom-right submatrix of $\mathcal{X}$, such that for each row of $X$, $r_i$ for $1 \leq i \leq M$, we have that $H \cdot r_i^T = S_i^T$, termed syndrome vector equation. This step is depicted in Supplementary Fig. 11c. Following this step, we can discard $S$, which will be reconstructed in the decoding process.

Note that after the last step, the first $M - r_3$ rows satisfy the required property $\mathbf{H} \cdot r_i^T = S_i^T$ by the definition of the vector S. Moreover, for the remaining $r_3$ rows, it is possible to solve this system of linear equations since **H** is a full-rank matrix with degree which is equal to $r_2 = 16$, and the last 16 columns are linearly independent.

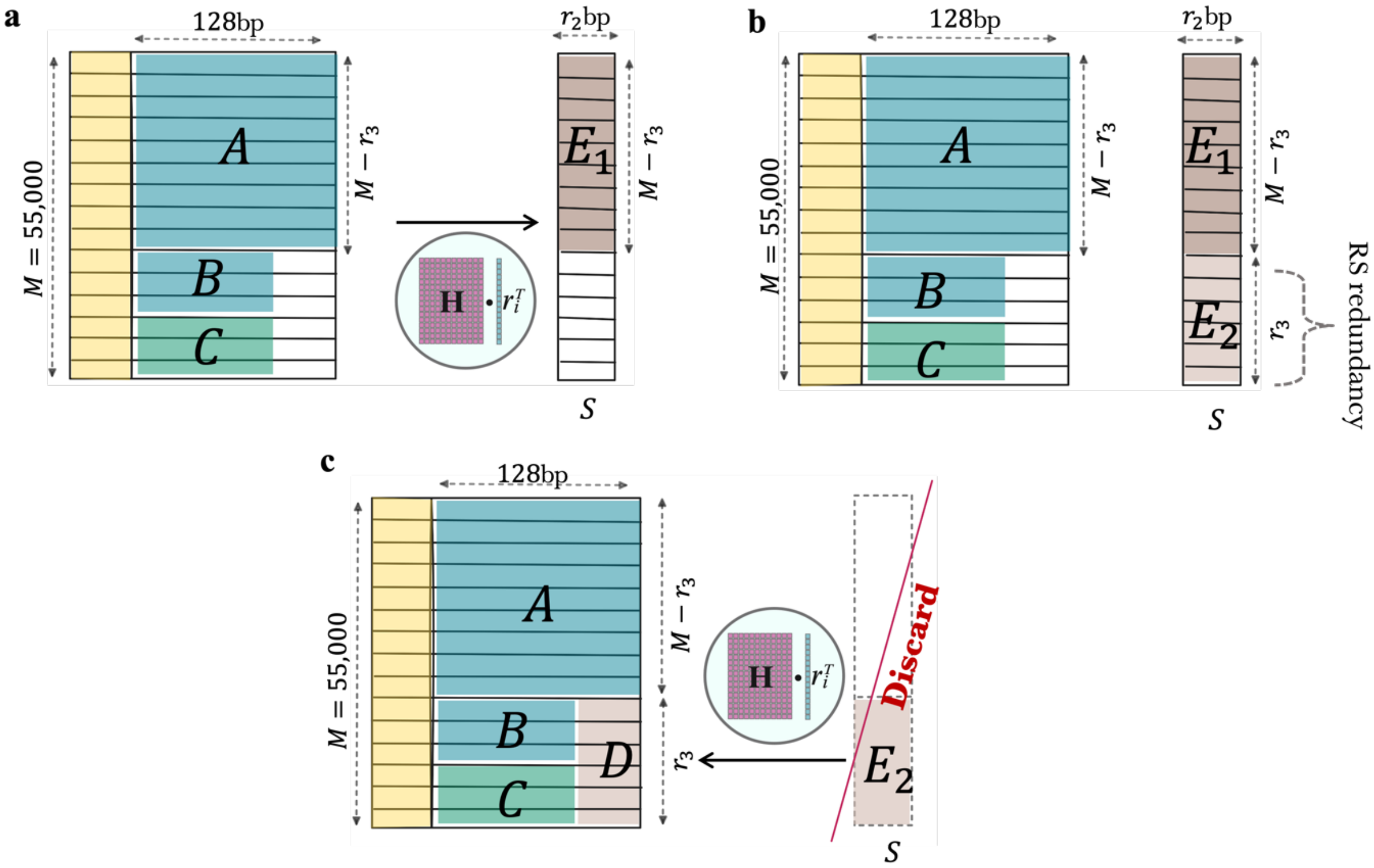

**Supplementary Fig. 11 | Description of the TP encoding. a,** multiple **H** by each row of $A$. **b,** encode E1 by an RS code. **c,** complete the matrix such that any row satisfies the syndrome vector equation. Following this step, we can discard $S$, which will be reconstructed in the decoding process.

## Decoding

We start by organizing the predictions from the DNAformer in a matrix $\hat{X}$, with $M$ rows, which is a noisy version of the matrix $X$. This can be done since our binning algorithm only considers correct indices. The matrix $\hat{X}$ is illustrated in Supplementary Fig. 12a. Let $\alpha$ denote the number of rows in $\hat{X}$ with three or less substitution errors, let $\beta$ denote the number of rows in $\hat{X}$ with more than three substitution errors, and let $\gamma$ denote the number of missing rows in $\hat{X}$. Given $\hat{X}$, our decoding process is based on the following steps.

1. **Recover the matrix $S$.** The first step in our decoding algorithm is to recover the matrix $S$ that was used during the TP encoding. To this end, we remove the 12 first symbols of each row in $\hat{X}$ (i.e. remove the index and file identifier of each row). Then we define $\widehat{S}$ to be the $M \times r_2$ matrix in which the i-th row, for $1 \leq i \leq M$, is equal to $\mathbf{H} \cdot r_i^T$, where $r_i$ is the i-th row of $\hat{X}$ (after removing the first 12 columns). Lastly, we decode the matrix $\hat{S}$ using $D_{RS}$, the decoder of the RS code that is used in our TP code. Note that for any i, if the prediction of DNAformer in the i-th row of $\hat{X}$ is correct, then the i-th row of $\hat{S}$ is equal to the i-th row of $S$. This implies that $\hat{S}$ is a noisy version of $S$, with at most $\alpha + \beta$ wrong rows and $\gamma$ missing rows. Hence, for any three positive integers $\alpha, \beta, \gamma$ such that $2(\alpha + \beta) + \gamma \leq r_3$, we have that $D_{RS}(\widehat{S}) = S$, i.e., $D_{RS}$ guarantees successful recover $S$ from $\hat{S}$. This process is illustrated in Supplementary Fig. 12b.

2. **Correct rows with up to 3 errors**. The next step is to correct the $\alpha$ rows in $\hat{X}$ that have 3 or less errors. This is done by first identifying the rows that are different in $S$ and $\hat{S}$. Then for each such row i, the i-th row of $\hat{X}$ is decoded using the parity-check matrix H and the i-th row of $S$ and $\widehat{S}$. Since H is a parity-check matrix of a BCH code that can correct up to 3 substitutions, each of the $\alpha$ rows with 3 or less errors will be corrected after this step, as depicted in Supplementary Fig. 12c.
3. **Recover the submatrices *A'*, *B'*, and *C'***. In this part of the decoding, we ignore the submatrix of $\hat{X}$ that corresponds to $D$ and only consider the submatrices $\hat{A}, \hat{B}$, and $\widehat{C}$, which are noisy versions of the submatrices $A, B$, and $C$. We first decode each 4-ary row (of the submatrices $\hat{A}, \hat{B}$, and $\widehat{C}$) into a row of bits using $D_{Cons}$, the decoder of our constrained code. Then we decode the columns of $\hat{A}', \widehat{B}'$, and $\widehat{C'}$ using $D_{DRS}$ the decoder of our diagonal RS code. Since any correct row in $\hat{A}, \hat{B}, \hat{C}$ will result in a correct binary row in $\hat{A}', \widehat{B}', \widehat{C'}$, respectively, these three submatrices together are a noisy version of the submatrix that is composed of $A', B'$, and $C'$, with at most $\beta$ wrong rows and $\gamma$ missing rows. Thus, for any two positive integers $\beta$ and $\gamma$ such that $2\beta + \gamma \leq r_1$, $D_{DRS}$ successfully recovers $A', B'$, and $C'$ from $\hat{A}', \widehat{B}', \hat{C}$. This step in presented in Supplementary Fig. 12d.
4. **Recover the submatrices *A*, *B*, and *C*.** To this end, we ignore the submatrix $\hat{A}''$ and convert $A', B'$, and $C'$ back to their 4-ary representation by encoding each row with $E_{cons}$ as presented in Supplementary Fig. 12e. Then, we use the matrix H again to recover the remaining symbols of $A$, which can be done since the $i$-th row of A, should satisfy $\mathbf{H} \cdot r_i^T = S_i^T$, where $S_i$ is the i-th row of the matrix S that we already recovered. This step is shown in Supplementary Fig. 12f.
5. **Recover the binary data $\boldsymbol{x}$**. Finally, as shown in Supplementary Fig. 12g, the submatrices A and B are converted back into bits using $D_{cons}$ and the binary data $\boldsymbol{x}$ is obtained by concatenating the rows of the latter.

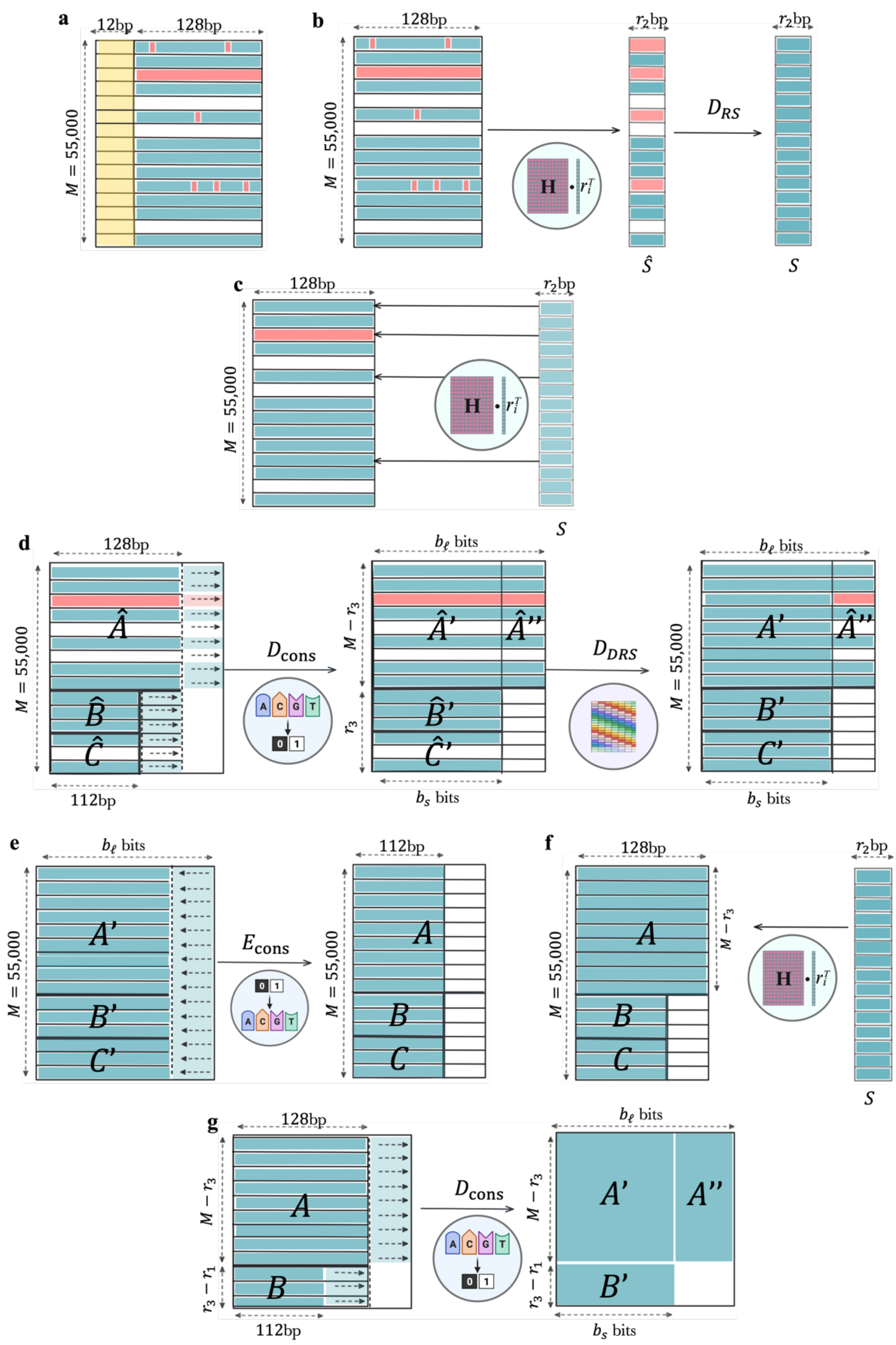

**Supplementary Fig. 12 | Description of the decoding step**. The colormap is as follows, the correct predications are the green rows. Substitutions are highlighted in red. There are $\alpha$ rows with three or less substitutions shown as green rows with few highlighted errors within them. There are $\beta$ rows with more than three errors and are shown as complete red rows. There are $\gamma$ rows which are the missing rows as shown as white rows. **a,** input to the decoder. **b,** reconstructing the syndrome vector $S$ using **H** matrix. **c,** using $S$ and **H** to reconstruct the $\alpha$ rows with up to three errors. **d,** transform the matrix back into binary format and use the diagonal encoder to correct *A', B'* and *C'*. **e,** transforming *A', B'*, and *C'* back to base 4. **f,** reconstruct $A$ using $S$ and **H**. **g,** transforming back into binary data.

## Analysis of decoder robustness

The discussion above implies that our decoding process recovers $X$ correctly if the following two conditions hold: $2(\alpha + \beta) + \gamma \leq r_3$, $2\beta + \gamma \leq r_1$. This relation is illustrated in Fig. 4b in which for any pair of parameters $\alpha$ and $\beta$ we present the maximal $\gamma$ for which our decoder is guaranteed to successfully decode $\hat{X}$ into $X$.

It can be observed that even for fixed values of $r_1, r_2$, and $r_3$, there is a wide range of values for $\alpha, \beta$, and $\gamma$, for which the decoding is successful. The robustness of our decoding algorithm is rooted in the tradeoff between these three quantities. More precisely, our confidence filter was designed to replace rows with substitutions (i.e., rows that were counted as either $\alpha$ or $\beta$) with missing rows (i.e., counted as $\gamma$). As $\alpha$ and $\beta$ have factor of two in our equations (and $\gamma$ doesn't), our confidence filter increases the safety margin of our entire DNA retrieval pipeline.

Additionally, when the CPL algorithm is activated, wrong prediction of the DNN can be corrected by the CPL reducing the total number of rows which are not correct in our pre-decoding matrix. In addition, the CPL can reduce the number of errors in the final DNAformer prediction which increase $\alpha$ while decreasing $\beta$. This improved the safety margin further as $r_3$ is larger than $r_1$.

The values of $\alpha, \beta$, and $\gamma$ for File 1 and File 2 are given in Supplementary Table 10, where the values are presented for the Illumina test dataset, and the two Nanopore test datasets. For each dataset, the three configurations of the DNAformer are considered. The first corresponds to the setup in which only the DNN is activated within the DNAformer. The second corresponds to the case in which the confidence filter is also activated. The last, which is our most robust setup, considers the configuration in which both the confidence filter and the CPL are activated within the DNAformer. The rows highlighted in red reflect cases where the decoder failed. The results show that for the Illumina data and the Nanopore two flowcells dataset, all the configurations of the DNAformer successfully retrieved the binary data. In the Nanopore single flowcell, which has the lowest SNR, the safety margin mechanism was needed to improve the DNAformer accuracy and retrieve the binary data successfully.

| | | Illumina test dataset | | | Nanopore test dataset single flowcell | | | Nanopore test dataset two flowcells | | |
|---|---|---|---|---|---|---|---|---|---|---|
| | | $\alpha$ | $\beta$ | $\gamma$ | $\alpha$ | $\beta$ | $\gamma$ | $\alpha$ | $\beta$ | $\gamma$ |
| File 1 | DNN | 2 | 1 | 28 | 1,337 | 1,440 | 34 | 882 | 479 | 14 |
| | DNN+Confidence | 2 | 1 | 28 | 1,229 | 827 | 814 | 879 | 437 | 65 |
| | DNN+Confidence+CPL | 2 | 1 | 28 | 1075 | 235 | 814 | 805 | 82 | 65 |
| File 2 | DNN | 2 | 1 | 28 | 1,351 | 1,422 | 38 | 843 | 515 | 10 |
| | DNN+Confidence | 2 | 1 | 28 | 1,273 | 841 | 779 | 836 | 477 | 57 |
| | DNN+Confidence+CPL | 2 | 1 | 28 | 1108 | 252 | 779 | 743 | 93 | 57 |

**Supplementary Table 10 | Analysis of the different types of errors for different DNAformer configurations.** The errors are shown for each of the three datasets and for each file separately. Red signifies failed retrieval. The results show that for the Illumina and Nanopore two flowcells, the DNN can retrieve the data on its own. However, for the Nanopore single flowcell the confidence filter and the CPL are needed to ensure retrieval.

## Confidence filter and safety margin

The safety margin is a metric that describes how robust the DNAformer is under specific working conditions. We derive the safety margin from the error correcting capabilities of the decoder shown in the previous section. It is defined as $\min(r_1 - 2\beta - \gamma, r_3 - 2(\alpha + \beta) - \gamma)$, where negative values correspond to cases in which the DNAformer fails to retrieve the information. We recall that, Supplementary Table 10 summarizes the values of $\alpha, \beta, \gamma$ for our Nanopore and Illumina datasets. In the table, the cases where the safety margin is negative are highlighted in red, in these cases the DNAformer fails to retrieve the information.

In our methodology, the design for a required safety margin is achievable via two control parameters, termed $cluster\ size_{threshold}$ which filters the minimum cluster size and $confidence_{threshold}$ which sends bad predictions of the DNN to the CPL.
The optimization process of the $cluster\ size_{threshold}$ included a similar methodology as described in the 'Effects of cluster size on the error rate' section. In this process we sampled different number of reads for each cluster from the Nanopore pilot dataset and examined the success rate, average Hamming distance and average edit distance of the DNN's predictions. This analysis resulted in $cluster\ size_{threshold} = 4$, where the success rate dropped below 50%. Similar behavior was observed in the validation analysis performed on the Nanopore two flowcells as shown in Supplementary Fig 1.

Based on our system methodology, the optimization process of the $confidence_{threshold}$ relies on an analysis of the pilot datasets. As the pilot datasets contains only 1,000 clusters, the average cluster size of the pilot dataset is much larger compared to the one in the test datasets which consists of 110,000 clusters. This gap in cluster size also creates a distribution which is similar in shape, however different in scale, as described in Supplementary Fig. 3. To overcome this gap and normalize the distributions, we sample 2% of the reads from the pilot dataset, and then perform the binning step. This process was repeated 110 times to create a normalized pilot dataset of up to 110,000 clusters, in which all the reads are taken from the real pilot datasets.

The optimization process of the $confidence_{threshold}$ is shown in Supplementary Fig. 13a where we show the safety margin for the normalized Nanopore pilot. In our design, we chose a safety margin of at least 1% or 1100 wrong predictions which corresponds to $confidence_{threshold} = 0.7$. Note that different values of safety margin will derive different values for the $confidence_{threshold}$. In cases where a very tight fit between the pilot and test dataset is required, we recommend a larger pilot dataset of 5,000-10,000 clusters.

Supplementary Fig. 13b shows the number of clusters sent to the CPL for various values of $confidence_{threshold}$. This graph illustrates the tradeoff between larger safety margin and the total runtime of the system. As more clusters are sent to the CPL, the overall runtime increases.

The test dataset shown in Supplementary Fig. 13a-b is the Nanopore single flowcell. This was chosen as the Illumina test dataset and Nanopore two flowcells test dataset are cases where the DNN can successfully retrieve the information without the confidence filter and CPL modules. However, the Nanopore single flowcell is a more challenging case which requires these modules for successful information retrieval.

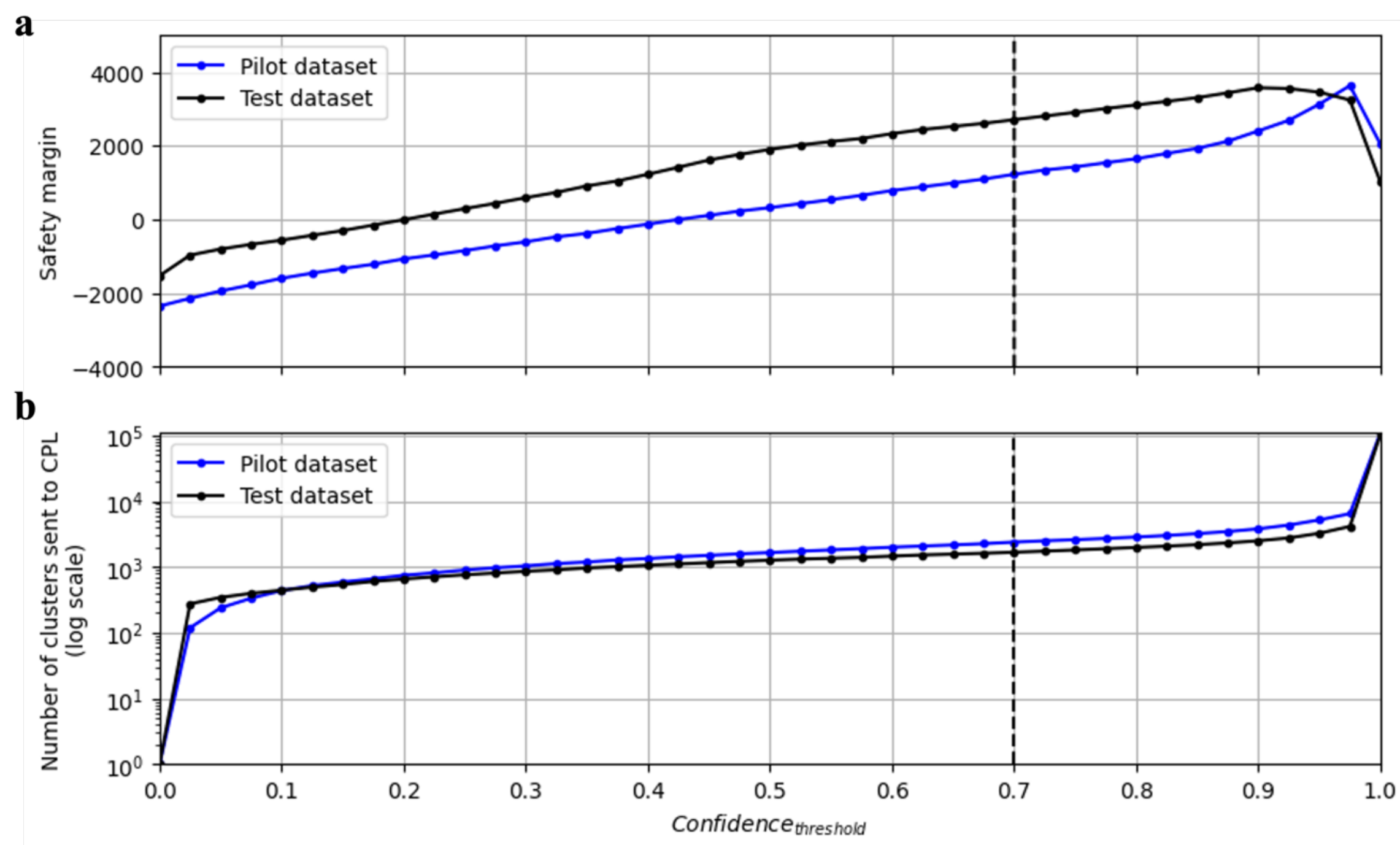

**Supplementary Fig. 13 | Analysis of the safety margin and confidence threshold.** Results of the confidence evaluation experiment on the Nanopore test single flowcell dataset. The X-axis represents the confidence threshold **a,** shows safety margin for the normalized Nanopore pilot and test datasets. In dashed line we show the chosen confidence threshold. **b,** shows the number of clusters that are sent to the CPL algorithm.

Lastly, we provide an analysis of the effect of the confidence function on our information retrieval pipeline. The analysis involved running both the DNN and CPL on the entire Nanopore test dataset single flowcell. Supplementary Fig. 14 presents Venn diagrams of the intersection between the number of clusters that have wrong DNN predictions, the number of clusters that were sent to the CPL algorithm by the confidence filter, and the number of clusters that have wrong CPL predictions. A wrong prediction is defined as an output from the DNN or CPL with at least one wrong symbol.

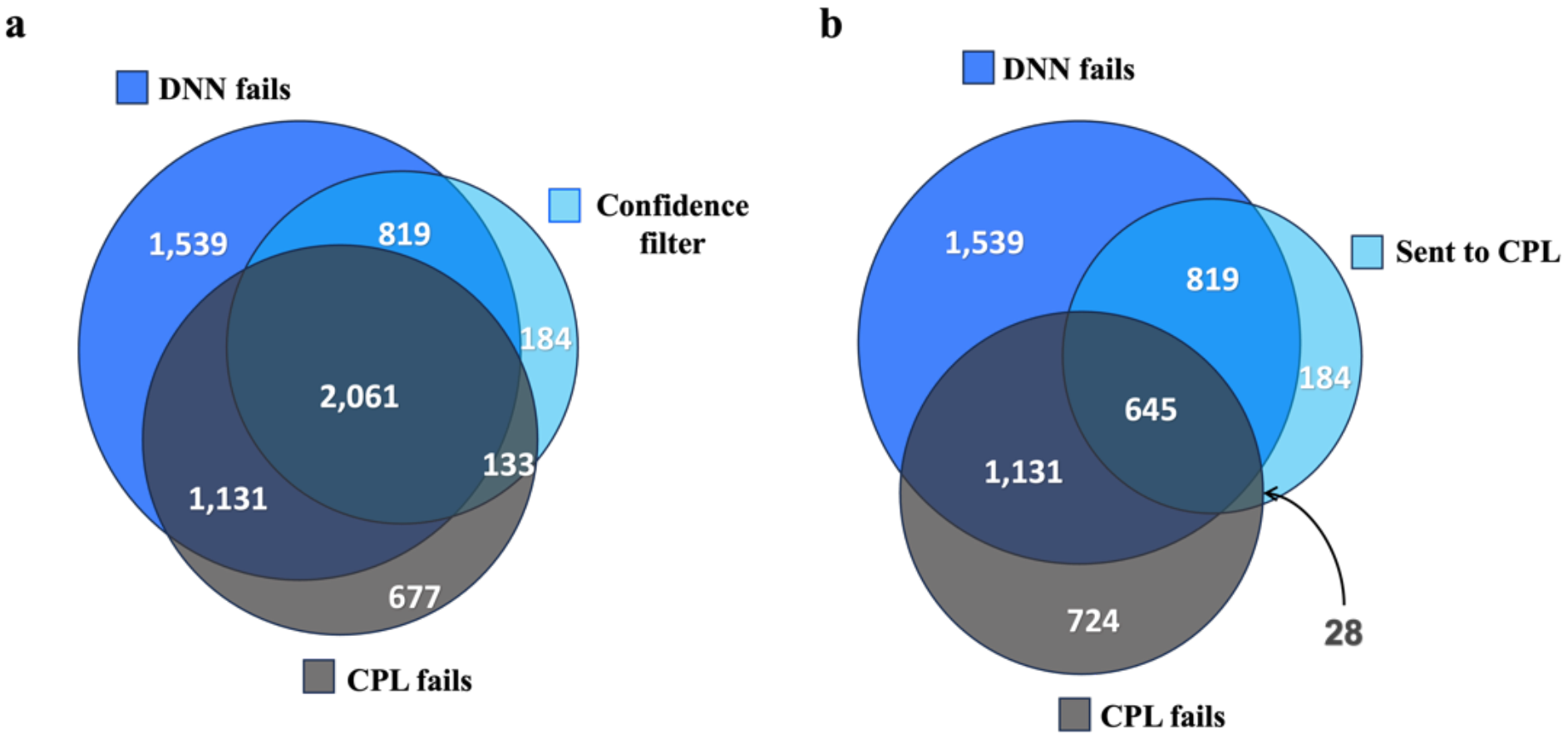

**Supplementary Fig. 14 | Detailed analysis of combining between the DNN, confidence filter and CPL.** Venn diagrams show the intersection between cases where the DNN fails, the CPL fails and the combination between them using the confidence filter to maximize performance. **a,** shows how the confidence filter interacts with the set of clusters for which the DNN or CPL fail. **b,** shows similar relations to **a** after filtering erasures.

The diagrams presented in Supplementary Fig. 14 show the benefits of incorporating the confidence filter and the CPL algorithm into our retrieval pipeline Supplementary Fig. 14a considers erasures as wrong predictions while Supplementary Fig. 14b ignores erasures. To fully understand the results, these two Venn diagrams should be considered together. The diagram in Supplementary Fig. 14a shows that among the 5,550 wrong predictions of the DNN, the confidence filter was able to detect 2,880, which are described in the Venn diagram by the intersection between DNN fails and Confidence filter. The 2,880 detected clusters comprise roughly 52% of these wrong predictions.
Note that the predictions filtered by the confidence filter correspond with three types of clusters: empty clusters, clusters that were passed to the CPL, and clusters that were screened by the confidence filter.
Shown in Supplementary Fig. 14b, among the DNN predictions, the number of clusters that were sent to the CPL is 1,676, which is 30% of the wrong and missing predictions of the DNN. Among these 1,676 clusters, 59% (1,003) were successfully corrected by CPL. In total, it means that out of all the wrong and missing predictions of the DNN, 15% were corrected by the CPL. Among the clusters that were sent to the CPL, there were 133 (4.1%), for which the DNN's outputs were correct while the CPL returned wrong predictions. For the rest of the clusters that were passed to the CPL, either the DNN and the CPL have corrected predictions, or both have wrong ones. Hence, they do not affect the final output.
To conclude, the diagrams show how incorporating the CPL into the pipeline increases the number of correct predictions and converts wrong predictions with missing ones, which increases the safety margin of our DNA retrieval pipeline.